\documentclass[aps,preprintnumbers,eqsecnum,amsmath,amssymb,showpacs,nofootinbib]{revtex4}
\usepackage{graphicx}
\usepackage{dcolumn}
\usepackage{bm}
\usepackage{epsfig}

\begin{document} 
\title{\boldmath Elastic and transition form factors of light pseudoscalar mesons from QCD sum rules}
\author{Irina Balakireva$^{1}$, Wolfgang Lucha$^{2}$, and Dmitri Melikhov$^{1,2,3}$}
\affiliation{$^1$SINP, Moscow State University, 119991, Moscow, Russia\\
$^2$HEPHY, Austrian Academy of Sciences, Nikolsdorfergasse 18, A-1050, Vienna, Austria\\
$^3$Faculty of Physics, University of Vienna, Boltzmanngasse 5, A-1090, Vienna, Austria}
\date{\today}
\begin{abstract}
We revisit $F_\pi(Q^2)$ and $F_{P\gamma}(Q^2)$, $P=\pi,\eta,\eta'$, making use of the local-duality (LD) 
version of QCD sum rules. We give arguments, that the LD sum rule provides reliable predictions for 
these form factors at $Q^2 \ge 5-6$ GeV$^2$, the accuracy of the method increasing 
with $Q^2$ in this region. For the pion elastic form factor, the well-measured data at small $Q^2$ give a hint 
that the LD limit may be reached already at relatively low values of momentum transfers, $Q^2\approx 4-8$ GeV$^2$; 
we therefore conclude that large deviations from LD in the region $Q^2=20-50$ GeV$^2$ seem very unlikely. 
The data on the ($\eta,\eta')\to\gamma\gamma^*$ form factors meet the expectations from the LD model. 
However, the {\sc BaBar} results for the $\pi^0\to\gamma\gamma^*$ form factor 
imply a violation of LD growing with $Q^2$ even at $Q^2\approx 40$ GeV$^2$, at odds with the $\eta,\eta'$ case 
and with the general properties expected for the LD sum rule.    
\end{abstract}
\maketitle

\section{Introduction}
In spite of the long history of theoretical investigations of the pion, its properties are still not 
fully understood. 

At asymptotically large values of the momentum transfer, $Q^2\to\infty$, the pion elastic and the $\pi\gamma$ transition 
form factors obey perturbative QCD (pQCD) factorization theorems 
\cite{pqcd,bl1980} 
\begin{eqnarray} 
\label{factorization}
F_{\pi}(Q^2) \to  8\pi\alpha_s(Q^2)f_\pi^2/Q^2,\qquad F_{\pi\gamma}(Q^2)\to  \sqrt{2} f_\pi/Q^2, 
\qquad  f_\pi=130 \mbox{ MeV}.   
\end{eqnarray} 
Subleading logarithmic and power corrections modify the behaviour (\ref{factorization}) at large but finite~$Q^2$. 

In early applications~of QCD to the pion elastic form factor, $F_\pi(Q^2)$, one hoped that power corrections vanish rather 
fast with $Q^2$; however, later investigations revealed that nonperturbative power corrections dominate~the form factor $F_\pi(Q^2)$ 
up to relatively high $Q^2\approx10$--$20$ GeV$^2$. This picture has arisen from different approaches 
\cite{isgur,simula,anisovich,roberts1,braun1,blm2008}. It was found that even at $Q^2$ as large as $Q^2=20$ GeV$^2$
the $O(1)$ term provides about half of~the form factor; the pQCD formula based on factorization 
starts to work well only at $Q^2\ge50$--$100$~GeV$^2$.

For the pion-photon transition form factor, $F_{\pi\gamma}(Q^2)$, axial anomaly \cite{abj,ab} fixes its value at $Q^2=0$. 
Brodsky and Lepage proposed a simple interpolating formula between the known value of the form factor at $Q^2=0$ and 
its asymptotic behaviour (\ref{factorization})
\begin{eqnarray}
\label{bl}
F_{\pi\gamma}(Q^2)=\frac{1}{2\sqrt{2}\pi^2f_\pi}
\left(1+\frac{Q^2}{4\pi^2f_\pi^2}\right)^{\!-1}, 
\end{eqnarray}
which was believed to describe well the transition form factor in a broad range of $Q^2$. 

\begin{figure}[ht!]
\begin{center}
\includegraphics[width=8.5cm]{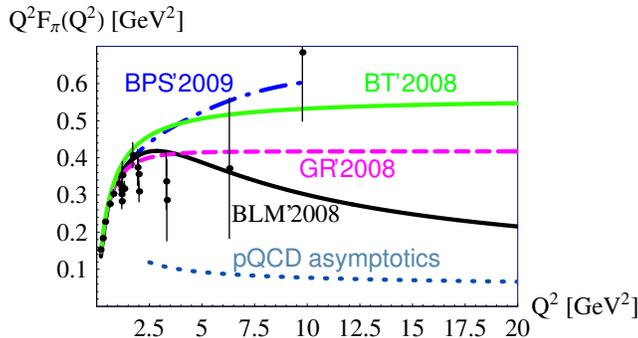}
\caption{\label{Plot:1}
Some predictions for the pion elastic form
factor $F_\pi(Q^2)$ --- lower solid (black) line:\ BLM'2008
\cite{blm2008}, upper solid (green) line:\ BT'2008 \cite{recent1},
dashed (magenta) line:\ GR'2008 \cite{recent2}, and dash-dotted
(blue) line:~BPS'2009 \cite[Eq.~(4.11b)]{bakulev} ---
vs.~experiment~\cite{data_piff}.}
\end{center}
\end{figure}

Surprisingly, some of the recent studies of the pion elastic form factor in the region  
$Q^2\approx 5-50$ GeV$^2$ \cite{recent1,recent2,recent3,bakulev} reported much larger valus of the 
pion elastic form factor than expected before (see Fig.~\ref{Plot:1}); 
the {\sc BaBar} result on the $\pi\to\gamma\gamma^*$ transition form factor \cite{babar} 
imply a strong violation of pQCD factorization in the region of $Q^2$ up to 40 GeV$^2$. 

In this paper, we revisit $F_\pi(Q^2)$ and $F_{P\gamma}(Q^2)$ making use of the local-duality (LD) version 
\cite{ld1,ld2} of QCD sum rules \cite{svz}. Our main emphasis is on the analysis of the expected 
accuracy of this powerful (although approximate) method for the calculation of hadron form factors. 

A local-duality sum rule is a dispersive three-point sum rule at $\tau=0$ 
(i.e., infinitely large Borel mass parameter). In this case all power corrections vanish 
and all details of the non-perturbative dynamics are hidden in one quantity -- 
the $Q^2$-dependent effective continuum threshold. Implementing duality in the standard way, i.e.,  
as the low-energy cut in the dispersion representations for the form factors, the sum rules 
relate the pion form factors to the low-energy region of Feynman diagrams of pQCD: 
\begin{eqnarray} 
\label{Fld1}
F^{\rm LD}_{\pi}(Q^2)= \frac{1}{f_\pi^2}\int\limits_0^{s_{\rm eff}(Q^2)}ds_1\int\limits_0^{s_{\rm eff}(Q^2)}ds_2 
\;\Delta_{\rm pert}(s_1, s_2, Q^2),\quad
\label{Fld2}
F^{\rm LD}_{\pi\gamma}(Q^2)=\frac{1}{f_\pi}\int\limits_0^{\bar s_{\rm eff}(Q^2)}ds\;\sigma_{\rm pert}(s, Q^2).    
\end{eqnarray} 
Here $\Delta_{\rm pert}(s_1,s_2,Q^2)$ is the double spectral density of the $\langle AVA\rangle$ 3-point function; 
whereas $\sigma_{\rm pert}(s,Q^2)$ is the single spectral density of the $\langle AVV \rangle$ 3-point function. 
These quantities are calculated as power series in $\alpha_s$: 
\begin{eqnarray} 
\label{densities}
\Delta_{\rm pert}(s_1, s_2, Q^2)=\Delta^{(0)}_{\rm pert}(s_1, s_2, Q^2)+
\alpha_s\Delta^{(1)}_{\rm pert}(s_1, s_2, Q^2)+O(\alpha_s^2), 
\quad \sigma_{\rm pert}(s, Q^2)=\sigma^{(0)}_{\rm pert}(s, Q^2)+O(\alpha_s^2).    
\end{eqnarray}
The one-loop spectral densities $\Delta^{(0)}_{\rm pert}$ and $\sigma^{(0)}_{\rm pert}$ are well-known 
\cite{m,teryaev1,ms}. 
The two-loop contribution $\Delta^{(1)}_{\rm pert}$ has been calculated in \cite{braguta}; 
the two-loop $O(\alpha_s)$ correction to $\sigma_{\rm pert}$ was found to be zero \cite{2loop}. 
Higher-order radiative corrections are unknown. 

If one knows the effective thresholds $s_{\rm eff}(Q^2)$ and $\bar s_{\rm eff}(Q^2)$, 
Eqs.~(\ref{Fld1}) provide the form factors. 
However, finding a reliable criterion for fixing the thresholds is a 
very subtle and difficult problem investigated in great detail in \cite{lms1}. 

Due to specific properties of the spectral functions at large values of the momentum transfer, 
the LD form factors (\ref{Fld1}) obey the 
factorization theorems (\ref{factorization}) as soon as the effective thresholds satisfy the following relations:  
\begin{eqnarray}
\label{ass}
s_{\rm eff}(Q^2\to\infty)=\bar s_{\rm eff}(Q^2\to\infty)={4\pi^2f_\pi^2}. 
\end{eqnarray}
For finite $Q^2$, however, the effective thresholds $s_{\rm eff}(Q^2)$ and $\bar s_{\rm eff}(Q^2)$ depend on $Q^2$ 
and differ from each other \cite{lms2}. The ``conventional LD model'' arises if one assumes (\ref{ass}) 
for all ``not too small'' values of $Q^2$ \cite{ld1}: 
\begin{eqnarray}
\label{sld}
s_{\rm eff}(Q^2)=\bar s_{\rm eff}(Q^2)={4\pi^2f_\pi^2}. 
\end{eqnarray}
Obviously, the LD model (\ref{sld}) for the effective continuum thresholds is an approximation which 
does not take into account details of the confinement dynamics. 
The only property of theory relevant for this model is factorization of hard form factors at large 
values of the monetum transfers. Because of the approximate character of the predictions from LD sum rules, 
it is important to understand the expected accuracy of the form factors obtained by this method. 
Quantum-mechanical potential models provide a possibility to probe this accuracy: one calculates the exact 
form factors by making use of the solutions of the Schr\"odinger equation and confronts these results with the 
application of LD sum rules in quantum mechanics \cite{qmsr,orsay,ms_inclusive}. 

This paper is organized as follows: 
In the next section, we recall details of LD sum rules in QCD and of the LD model for form factors. 
Section 3 studies the accuracy of the LD model for elastic and transition form factors in a 
quantum-mechanical potential model. 
The pion elastic form factor is discussed in Section 4 and the $P\to\gamma\gamma^*$ transition from factors 
are considered in Section 5. Section 6 gives our conclusions. 
Appendix A provides technical details of perturbative two-loop calculations in non-relativistic field theory.

\section{Local-duality model for form factors in QCD}

The sum-rule calculations of the hadron form factors are based on the OPE for the relevant correlation functions, 
which contain the contributions of ground-state hadrons of interest. In order to extract the form factors of 
pseudoscalar mesons, one analyses the $\langle AVA\rangle$ and $\langle AVV\rangle$ correlators, $A$ 
being the axial and $V$ the vector currents. 

\subsection{\label{Sec:2}The three-point function \boldmath $\langle AVA\rangle$ and the pion elastic form factor}    
The basic objects for the extraction of the pion decay constant and the elastic form factor are the 
two- and three-point correlation functions
\begin{eqnarray}
\label{sr_pion}
\Pi\!\left(p^2\right)&=&\int{\rm d}^4x\,e^{{\rm i}px}
\left<0\left|T[j(x)\,j^\dag(0)]\right|0\right>,
\nonumber
\\ 
\Gamma\!\left(p^2_1,p^2_2,q^2\right)&=&\int{\rm d}^4x_1\,{\rm d}^4x_2\,e^{{\rm i}(p_1x_1-p_2x_2)} 
\left<0\left|T[j(x_1)\,J(0)\,j^\dag(x_2)]\right|0\right>,
\qquad q\equiv p_1-p_2,\qquad Q^2\equiv-q^2;
\end{eqnarray}
where $|0\rangle$ is the physical QCD vacuum, which differs from pQCD vacuum; 
properties of the physical QCD vacuum are characterized by the condensates \cite{svz}. 
$j$ is shorthand for the interpolating axial current $j_{5\alpha}$ of the positively charged pion, 
\begin{eqnarray}
\left<0\left|j_{5\alpha}(0)\right|\pi(p)\right>={\rm i}p_{\alpha}f_{\pi}. 
\end{eqnarray}
$J$ labels the electromagnetic current
$J_{\mu}$, and for brevity we omit all Lorentz indices. In QCD,
the correlators (\ref{sr_pion}) can be found by applying OPE. Instead of discussing the Green
functions~(\ref{sr_pion})~in~Minkowski space, it is convenient to
study the time-evolution operators in Euclidean space, which arise
upon performing the Borel transformation $p^2\to\tau$ to a
parameter $\tau$ related to Euclidean time. The Borel image of the
two-point correlator~$\Pi(p^2)$~is
\begin{eqnarray}
\label{SR_P2}
\Pi_{\rm OPE}(\tau)=\int\limits_0^\infty{\rm d}s\,e^{-s\tau}\,\rho_{\rm pert}(s)+\Pi_{\rm cond}(\tau),\qquad\rho_{\rm pert}(s)=
\rho_0(s)+\alpha_s\,\rho_1(s)+O(\alpha^2_s),\end{eqnarray}with
spectral densities $\rho_i(s)$ related to perturbative two-point
graphs, and nonperturbative power corrections $\Pi_{\rm cond}(\tau)$. At hadron level, insertion of intermediate hadron
states casts the Borel-transformed two-point correlator into the form
\begin{equation}
\label{SR_P3}
\Pi(\tau)=f_\pi^2\,e^{-m_\pi^2\tau}+\mbox{excited states}.
\end{equation}
In this expression for $\Pi(\tau),$ the first term on the right-hand side constitutes the pion contribution. 
Applying the~double Borel transform $p^2_{1,2}\to\tau/2$ to the three-point correlator
$\Gamma(p^2_1,p^2_2,q^2)$ results, at QCD level, in
\begin{align}
\label{3PF}
\Gamma_{\rm OPE}(\tau,Q^2)&=
\int\limits_0^\infty\,\int\limits_0^\infty{\rm d}s_1\,{\rm d}s_2 \exp\!\left(-\frac{s_1+s_2}{2}\,\tau\right)
\Delta_{\rm pert}(s_1,s_2,Q^2)+\Gamma_{\rm cond}(\tau,Q^2),\nonumber\\
\Delta_{\rm pert}(s_1,s_2,Q^2)&=\Delta_0(s_1,s_2,Q^2)
+\alpha_s\,\Delta_1(s_1,s_2,Q^2)+O(\alpha^2_s),
\end{align}
where
$\Delta_{\rm pert}(s_1,s_2,Q^2)$ is the double spectral density of
the three-point graphs of perturbation theory and $\Gamma_{\rm
cond}(\tau,Q^2)$ labels the power corrections. Inserting
intermediate hadron states yields, for the hadron-level
expression~for~$\Gamma(\tau,Q^2),$
\begin{equation}
\Gamma(\tau,Q^2)=F_\pi(Q^2)\,f_\pi^2\,e^{-m_\pi^2\tau}
+\mbox{excited states}.
\end{equation}
Quark--hadron duality assumes that above effective
continuum thresholds $s_{\rm eff}$ the excited-state contributions
are~dual to the high-energy regions of the perturbative graphs. In
this case, the relevant sum rules read in the chiral limit \cite{ld1,ioffe}
\begin{eqnarray}
\label{SR_P4}
f_\pi^2&=&\int\limits_0^{\tilde s_{\rm eff}(\tau)}{\rm d}s\, e^{-s\tau}\,\rho_{\rm
pert}(s)+\frac{\left<\alpha_s\,G^2\right>}{12\pi}\,\tau
+\frac{176\pi\,\alpha_s\left<\bar qq\right>^2}{81}\,\tau^2+\cdots,\\ 
\label{SR_G4}
F_\pi(Q^2)\,f_\pi^2&=&\int\limits_0^{s_{\rm
eff}(Q^2,\tau)}\,\int\limits_0^{s_{\rm eff}(Q^2,\tau)}{\rm d}s_1\,{\rm d}s_2\, \Delta_{\rm pert}(s_1,s_2,Q^2)
\exp\!\left(-\frac{s_1+s_2}{2}\,\tau\right)\nonumber\\
&&+\frac{\left<\alpha_s\,G^2\right>}{24\pi}\,\tau
+\frac{4\pi\,\alpha_s\left<\bar qq\right>^2}{81}\,\tau^2
\left(13+Q^2\,\tau\right)+\cdots.
\end{eqnarray}
As a consequence of the use of local condensates, the right-hand
side of (\ref{SR_G4}) involves polynomials in $Q^2$ and~therefore
increases with $Q^2$, whereas the form factor $F_\pi(Q^2)$ on the
left-hand side should decrease with $Q^2.$ Hence, at large
$Q^2$ the sum rule (\ref{SR_G4}), with its truncated series of
power corrections, cannot be directly used. There are
essentially two ways for considering the region of large
$Q^2$.

One remedy is the resummation of power corrections: the
resummed power corrections decrease with increasing $Q^2.$ This
may be achieved by the introduction of nonlocal condensates
\cite{nonlocal} in a, however, model-dependent
manner~\cite{chernyak_2006}.

Another option is to fix the Borel parameter
$\tau$ to the value $\tau=0,$ thus arriving at a 
local-duality sum~rule \cite{ld1,ld2}. Therein all power
corrections vanish and the remaining perturbative term decreases
with $Q^2$. In the~LD limit, one finds
\begin{eqnarray}
\label{SR_P5}
f_\pi^2&=&\int\limits_0^{\tilde s_{\rm eff}}{\rm d}s\,\rho_{\rm pert}(s)=\frac{\tilde s_{\rm eff}}{4\pi^2}
\left(1+\frac{\alpha_s}{\pi}\right)+O\!\left(\alpha_s^2\right),
\\
\label{SR_G5}
F_\pi(Q^2)\,f_\pi^2&=&\int\limits_0^{s_{\rm eff}(Q^2)}\,
\int\limits_0^{s_{\rm eff}(Q^2)}{\rm d}s_1\,{\rm d}s_2\, \Delta_{\rm pert}(s_1,s_2,Q^2).
\end{eqnarray}
The spectral densities $\rho_{\rm pert}(s)$ and $\Delta_{\rm pert}(s_1,s_2,Q^2)$ are calculable by perturbation theory. 
Hence, by fixing $\tilde s_{\rm eff}$ and $s_{\rm eff}(Q^2)$, it is
straightforward to extract the pion's decay constant $f_\pi$ and
form factor $F_\pi(Q^2).$

Noteworthy, the effective and the physical thresholds are different quantities: The latter is a constant determined by the
masses~of the hadron states. The effective thresholds $\tilde s_{\rm eff}$ and $s_{\rm eff}$ are parameters of the sum-rule
method~related~to the specific realization of quark--hadron
duality; in general, they are {\em not\/} constant but depend on external kinematical variables \cite{lms1,lms2}.

Let us recall the important properties of the spectral densities
on the right-hand sides of (\ref{SR_P2}) and (\ref{3PF}):
For~$Q^2\to0,$ the Ward identity relates the spectral densities
$\rho_i(s)$ and $\Delta_i(s_1,s_2,Q^2)$ of two- and three-point
functions~to~each~other:
\begin{eqnarray}
\label{Ward_identity}
\lim_{Q^2\to0}\Delta_i(s_1,s_2,Q^2)=\rho_i(s_1)\,\delta(s_1-s_2),\qquad
i=0,1,\dots.
\end{eqnarray}
For $Q^2\to\infty$ and $s_{1,2}$ kept
fixed, explicit calculations \cite{blm2008} yield
\begin{eqnarray}
\label{large_momentum}\lim_{Q^2\to\infty}\Delta_0(s_1,s_2,Q^2)\propto
\frac{1}{Q^4},\qquad\lim_{Q^2\to\infty}\Delta_1(s_1,s_2,Q^2)
=\frac{8\pi}{Q^2}\,\rho_0(s_1)\,\rho_0(s_2).
\end{eqnarray}
For the pion form factor $F_\pi(Q^2)$ on the left-hand side of
(\ref{SR_G4}) two exact properties are known, namely, its
normalization condition related to current conservation, requiring
$F_\pi(0)=1,$ and the factorization theorem (\ref{factorization}).
Obviously, for 
\begin{equation}
\label{seff-sL} 
s_{\rm eff}(Q^2\to0)=\frac{4\pi^2\,f^2_{\pi}}{1+\alpha_s(0)/\pi},\qquad
s_{\rm eff}(Q^2\to\infty)=s_{\rm LD}\equiv4\pi^2\,f_\pi^2,
\end{equation}
the form factor $F_\pi(Q^2)$ extracted from the LD
sum rule (\ref{SR_G5}) satisfies both of these rigorous
constraints. At small~$Q^2,$ we assume a freezing of
$\alpha_s(Q^2)$ at the level 0.3, as is frequently done.
At the intermediate $Q^2$, the effective threshold depends on $Q^2$ and the pion form factor obtained from the LD 
sum rule depends on the details of $s_{\rm eff}(Q^2)$. The ``conventional LD model'' {\it assumes} that reasonable predictions 
for the elastic form factor at not too small values of $Q^2$ may be obtained by setting $s_{\rm eff}(Q^2)=4\pi^2f_\pi^2$, 
see Eq.~(\ref{sld}). 

Although one has to invoke assumptions on behaviour of the effective threshold at the intermediate $Q^2$, 
some features of the pion form factor turn out to be largely independent of this assumption. 
So, let us look more carefully what is in fact conjectured in the LD sum rule and what may be predicted by this approach. 

\begin{figure}[ht!]
\begin{center}
\includegraphics[width=8.5cm]{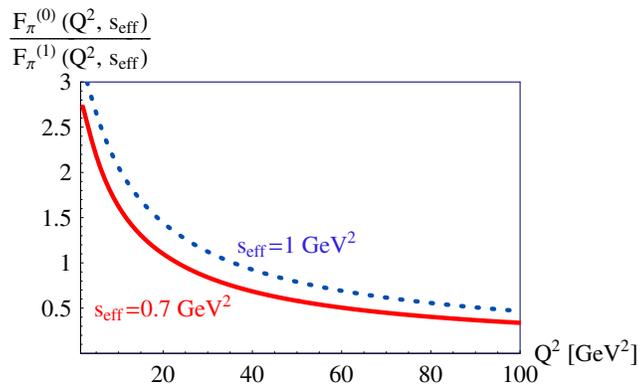}
\caption{\label{Plot:1d}
Ratio $F_\pi^{(0)}(Q^2)/F_\pi^{(1)}(Q^2)$
of $O(1)$ and $O(\alpha_s)$ contributions to the pion elastic form
factor vs.~$Q^2$ for two reasonable~values of the --- here by
assumption constant --- effective threshold $s_{\rm eff}$.}
\end{center}
\end{figure}

The sum rule (\ref{SR_G5}) for the pion form factor relies on two
ingredients: first, on the {\it rigorous calculation\/} of the
spectral densities of the perturbative-QCD diagrams (recall that
power corrections vanish in the LD limit $\tau=0$); second, on~the
{\it assumption\/} of quark--hadron duality, which claims that the
contributions of the hadronic continuum states may be well 
described by the diagrams of perturbation theory above some
effective threshold $s_{\rm eff}$. Thus, the only ---
although really essential --- unknown ingredient of the LD sum rule for
the pion elastic from factor is this effective continuum threshold
$s_{\rm eff}(Q^2).$ Let us emphasize that, since the $O(1)$ and
$O(\alpha_s)$ contributions to the pion form factor are governed
by one and the same effective threshold $s_{\rm eff}(Q^2)$, the
relative weights of these contributions may be {\it predicted}.
Their~ratio~$F_\pi^{(0)}(Q^2)/F_\pi^{(1)}(Q^2)$ turns out to be
relatively stable with respect to $s_{\rm eff}$ and may therefore
be calculated relatively accurately (see Fig.~\ref{Plot:1d}).

Quark--hadron duality implies that the effective threshold
(although being a function of $Q^2$) always --- i.e.,~also for
large $Q^2$ --- stays in a region close to $1$ GeV$^2.$ Moreover,
in order to satisfy the QCD factorization theorem for~the
$O(\alpha_s)$ contribution to the form factor, the effective
threshold should behave like $s_{\rm eff}(Q^2)\to4\pi^2f_\pi^2$
for $Q^2\to\infty.$ This requirement has immediate consequences
for the large-$Q^2$ behaviour of different contributions to the
pion form factor:

\noindent 1. Since $s_{\rm eff}(Q^2)$
is bounded from above, the $O(1)$ contribution $F_\pi^{(0)}(Q^2)$
to the elastic form factor $F_\pi(Q^2)$ of~the~pion behaves like
$F_\pi^{(0)}(Q^2)\propto1/Q^4$ for $Q^2\to\infty.$ We would like
to emphasize that the decrease of the soft contribution to the
pion elastic form factor like $1/Q^4$ is a~direct consequence of
perturbation theory and quark--hadron duality.

\noindent 2. Consequently,
for large $Q^2$ the pion elastic form factor $F_\pi(Q^2)$ is
dominated by the $O(\alpha_s)$ contribution $F_\pi^{(1)}(Q^2).$

Now, in the holographic models of \cite{recent1,recent2} merely
the soft contribution is considered: it behaves like $1/Q^2$ for
large~$Q^2$. This is only possible if the effective threshold
$s_{\rm eff}(Q^2)$ rises with $Q^2.$ However, this immediately
leads to the violation of the factorization theorem for the
$O(\alpha_s)$ contribution, which is governed by the same
effective threshold. Consequently, we would like to emphasize that
the findings of \cite{recent1,recent2} would imply that the QCD
factorization theorem is violated.~~We thus conclude that the
predictions of \cite{recent1,recent2} for the pion form factor
seem to us improbable.\footnote{A way out would be to assume
different effective thresholds for the $O(1)$ and the
$O(\alpha_s)$ contributions to the pion form factor. This~seems,
however, a rather artificial construction.}

\subsection{The three-point function {\boldmath $\langle VAV\rangle$} and the $P\to\gamma$ transition form factor}
Let us now consider the amplitude of two-photon production from the
vacuum induced by the axial-vector current~of nearly massless
quarks of one flavour, $j_\mu^5=\bar q\gamma_\mu\gamma_5q$, 
with $\varepsilon_{1,2}$ denoting the photon polarization vectors:
\begin{equation}
\langle\gamma(q_1)\gamma(q_2)|j_\mu^5(x=0)|0\rangle=
T_{\mu\alpha\beta}(p|q_1,q_2)\varepsilon^\alpha_1\varepsilon^\beta_2,
\qquad p=q_1+q_2.
\end{equation}
The amplitude $T_{\mu\alpha\beta}$
is obtained from the vacuum expectation value of the $T$-product
of two vector and one axial-vector currents and will be referred to
as the $\langle VAV\rangle$ amplitude. Vector-current conservation
yields the following relations:
\begin{eqnarray}
T_{\mu\alpha\beta}(p|q_1,q_2)q_1^\alpha=0,\qquad
T_{\mu\alpha\beta}(p|q_1,q_2)q_2^\beta=0.
\end{eqnarray}
The general decomposition of the amplitude may be written as
\begin{eqnarray}
\label{amp2}
T_{\mu\alpha\beta}(p|q_1,q_2)=-p_\mu \epsilon_{\alpha\beta q_1 q_2}i F_0 
+(q_1^2\epsilon_{\mu\alpha\beta q_2}-q_{1\alpha}\epsilon_{\mu q_1\beta q_2})i F_1
+(q_2^2\epsilon_{\mu\beta\alpha q_1}-q_{2\beta} \epsilon_{\mu q_2\alpha q_1})i F_2.
\end{eqnarray}
The form factor $F_0$ contains the contribution of the pseudoscalar meson of our interest \cite{lm2011} such that 
\begin{eqnarray}
F_{P\to\gamma\gamma}(q_1^2,q_2^2)=-\frac{1}{f_P}F_0(p^2=m_P^2|q_1^2,q_2^2).
\end{eqnarray} 
We also consider the transition amplitude of the pseudoscalar current operator $\bar q\gamma_5q$:
\begin{eqnarray}\label{G5}
\langle\gamma(q_1)\gamma(q_2)|\bar q\gamma_5q|0\rangle=
\epsilon_{\alpha\beta q_1q_2}\varepsilon^\alpha_1\varepsilon^\beta_2 F_5(q_1^2,q_2^2,p^2).
\end{eqnarray}
The two-photon amplitude of the divergence of the axial current takes the form
\begin{eqnarray}
\label{anomaly1}
\langle\gamma(q_1)\gamma(q_2)|\partial^\mu j^5_\mu|0\rangle=
\epsilon_{\alpha\beta q_1 q_2}\varepsilon^\alpha_1 
\varepsilon^\beta_2 (p^2 F_0-q_1^2 F_1-q_2^2F_2).
\end{eqnarray}
The case of our interest is $q_1^2=0$, then the form factor $F_1$
does not contribute to the divergence. In perturbation theory, the
form factors $F_0$, $F_2$, and $F_5$ may be written in terms of
their spectral representations in $p^2$ (with $q^2\equiv q_2^2=-Q^2$):
\begin{eqnarray}
F_i(p^2,q^2)=\frac{1}{\pi}
\int\limits_{4m^2}^\infty\frac{ds}{s-p^2}\,\Delta_i(s,q^2).
\end{eqnarray}
To one-loop order, the spectral densities read \cite{ms,m,teryaev1}
\begin{eqnarray}
\label{1loop}
\Delta_0(s,q^2)&=&-\frac{1}{2\pi}\frac{1}{(s-q^2)^2}
\left[-q^2\,w+2m^2\log\left(\frac{1+w}{1-w}\right)\right],\nonumber\\
\Delta_2(s,q^2)&=&-\frac{1}{2\pi}\frac{1}{(s-q^2)^2}
\left[-s\,w+2m^2\log\left(\frac{1+w}{1-w}\right)\right],\nonumber\\
\Delta_5(s,q^2)&=&-\frac{1}{2\pi}\frac{m}{s-q^2}
\log\left(\frac{1+w}{1-w}\right),\qquad w\equiv\sqrt{1-4m^2/s}.
\end{eqnarray}
Obviously, the absorptive parts $\Delta_i$ obey the classical
equation of motion for the divergence of the axial current
\begin{eqnarray}
s\,\Delta_0(s,q^2)-q^2\,\Delta_2(s,q^2)=2m\,\Delta_5(s,q^2).
\end{eqnarray}
The form factors then satisfy
\begin{eqnarray}
\label{3.7}
p^2F_0(p^2,q^2)-q^2F_2(p^2,q^2)=2m\,F_5(p^2,q^2)
-\frac{1}{\pi}\int\limits_{4m^2}^\infty ds \,\Delta_0(s,q^2).
\end{eqnarray}
The last integral is equal to $-{1}/{2\pi}$, independently of the values of
$m$ and $q^2$, and represents the axial anomaly \cite{abj}:
\begin{eqnarray}
\label{dd}
p^2F_0(p^2,q^2)-q^2F_2(p^2,q^2)=2m\,F_5(p^2,q^2)+\frac{1}{2\pi^2}.
\end{eqnarray}
In the chiral limit $m=0$ and for $q^2=0$, the form factor $F_0$
develops a pole related to a massless pseudoscalar~meson \cite{zakharov}. 
The residue of this pole is again the axial-anomaly $1/{2\pi^2}$. 
Notice however that the pole in the one-loop expression for $F_0$ disappears as soon as 
one of the photons is virtual \cite{zakharov,teryaev1,ioffe2}. 

As is clear from (\ref{3.7}), the anomaly represents the integral of $\Delta_0$, 
the spectral density of the form factor $F_0$. Adler and Bardeen \cite{ab} tell us that the anomaly is 
non-renormalized by multiloop corrections. The easiest realization of this property would have been just 
the vanishing of multiloop contributions to the spectral density $\Delta_0(s,q^2)$. An argument in favour of 
this possibility comes from explicit two-loop calculations \cite{2loop} which report the  
non-renormalizability of the full $\langle VAV\rangle$ vertex to the two-loop accuracy. 
However, if so, the full form factor $F_0$ is given by its one-loop expression. Then, this expression should 
develop the pion pole, known to be present in the full amplitude for any value of $q^2$. 
But, obviously, this pole does not emerge in the one-loop expression for $F_0$ if $q^2\ne 0$! 

This requires that multiloop corrections to the form factor $F_0$ (and, respectively, to its absorptive part
$\Delta_0(s,q^2)$) do not vanish. Then one may ask oneself how it may happen that the anomaly nevertheless remains 
non-renormalized by multiloop corrections? The only possible answer we see \cite{lm2011} is that the non-renormalization 
of the anomaly is reached due to some conspiral property of multiloop contributions to $\Delta_0(s,q^2)$ 
forcing their integral to vanish 
(in fact, quite similar to the one-loop result: although the spectral density explicitly depends on $m$ and $q^2$, 
its integral is an $m$- and $q^2$-independent constant). 

As is obvious from (\ref{amp2}), the contribution of the light pseudoscalar constitutes a part of the form factor $F_0$. 
The Borel sum rule for the corresponding Lorentz structure reads ($Q^2=-q^2>0$)
\begin{eqnarray}
\int ds\exp(-s\tau)\Delta_0(s,Q^2)=-f_P F_{P\gamma}(Q^2)\exp(-m_P^2\tau)+\mbox{contributions of excited states}.
\end{eqnarray}
Exploiting the concept of duality, the contribution of the excited
states is assumed to be dual to the high-energy region of the
diagrams of perturbation theory above an effective threshold
$s_{\rm eff}$. After that, setting the Borel parameter $\tau=0$
we arrive at the
LD sum rule for a pseudoscalar $\bar q q$-meson 
\begin{eqnarray}
\label{ld1}
\int\limits_{4m^2}^{s_{\rm eff}(Q^2)}ds\,\Delta_0(s,Q^2)=- f_P F_{P\gamma}(Q^2).
\end{eqnarray}
The spectral density $\Delta_0$ to one-loop order is given by (\ref{1loop}); two-loop corrections were 
found to be absent \cite{2loop}. As discussed above, higher-loop corrections to $\Delta_0$ 
cannot vanish; so the l.h.s. of 
(\ref{ld1}) is known to $O(\alpha_s^2)$ accuracy. 

In the chiral limit, the LD expression for the form factor for the one-flavour case is particularly simple (cf. \cite{ld2,teryaev2}):
\begin{eqnarray}
\label{ld}
F_{P\gamma}(Q^2)=\frac{1}{2\pi^2f_P}\frac{s_{\rm eff}(Q^2)}{s_{\rm eff}(Q^2)+Q^2}.
\end{eqnarray}
We would like to emphasize that the effective threshold in (\ref{ld}) depends on $Q^2$. 
Apart from neglecting $\alpha_s^2$ and higher-order corrections to the spectral density $\Delta_0$, 
no approximations have been done up to now: we have just considered the LD limit $\tau=0$; 
for an appropriate choice of $s_{\rm eff}(Q^2)$ the form factor may still be calculated exactly. 
Approximations come into the game when we consider a model for $s_{\rm eff}(Q^2)$.

Independently of the behaviour of $s_{\rm eff}(Q^2)$, the form factor at $Q^2=0$ is related to the axial anomaly: 
$F_{P\gamma}(0)=1/(2\pi^2f_P)$. QCD factorization requires $s_{\rm eff}(Q^2)\to 4\pi^2f_P^2$ for large $Q^2$. 
So, the simplest model compatible with this requirement is obtained by setting $s_{\rm eff}(Q^2)= 4\pi^2f_P^2$ 
for {\it all} values of $Q^2$ \cite{ld1}. This is the ``conventional LD model'' (\ref{sld}) yielding for the neutral 
pion case the Brodsky--Lepage interpolating formula \ref{bl}.\footnote{In an alternative 
approach to the $P\gamma$ form factor \cite{agaev,mikhailov},  
the pseudoscalar meson is described by a set of distribution amplitudes of increasing twist which are 
treated as nonperturbative inputs. 
In our analysis, the deviation of the effective threshold $s_{\rm eff}(Q^2)$ from its asymptotic value $4\pi^2 f_P^2$ 
corresponds to some extent to the contribution of higher-twist distribution amplitudes in the 
approach of \cite{mikhailov}.}

As already pointed out above, the conventional LD model for
the effective continuum threshold (\ref{sld}), defined by
the~choice $s_{\rm eff}(Q^2)=\bar s_{\rm eff}(Q^2)=4\pi^2f_\pi^2$ for all $Q^2,$ or a 
slightly more sophisticated approach of \cite{blm2008} are approximations which do not 
account for subtle details of
the confinement dynamics. Consequently, it is important 
to understand the accuracy to be expected within this approach; in other words, 
it is important to obtain some reliable estimate~of the expected deviations of the 
exact $s_{\rm eff}(Q^2)$ from its LD limit $s_{\rm LD}$ in the momentum region $Q^2\ge 4$--$6$ GeV$^2$. 
To this end, in the next Section we take advantage of the
fact that in quantum mechanics all the bound-state properties may be 
found~exactly~by solving the~Schr\"odinger~equation. On the other
hand, also in quantum mechanics we may construct LD sum rules.


\section{Exact vs. LD form factors in quantum-mechanical potential models}

The relevance and the expected accuracy of the LD model may be tested in those cases where the form factor 
$F_{P\gamma}(Q^2)$ is known, i.e., may be calculated by other theoretical approaches or measured
experimentally. Then, the exact effective~threshold may be reconstructed from (\ref{ld}), 
in this way probing the accuracy of the LD model. 

To probe the accuracy of the LD model, we now consider a quantum-mechanical example: 
the corresponding form factors may be calculated using the solution of the Schr\"odinger equation 
and confronted with the results of the quantum-mechanical LD model, which is constructed 
precisely the same way as in QCD. For the elastic form factor, it is mandatory 
to consider a potential involving both the Coulomb and the confining parts; 
for the analysis of the transition form factor one may start with a purely confining potential. 

The basic object for quantum-mechanical LD sum rules is the analogue of the three-point correlator of field theory \cite{lms2}
\begin{eqnarray}
\label{3pointQM}
\Gamma^{\rm NR}(E,E',Q)=\langle r'=0|\frac{1}{H-E'} J(\bm{q})\frac{1}{H-E} |r=0 \rangle, 
\qquad Q\equiv |\bm{q}|. 
\end{eqnarray}
Here, $H$ is the Hamiltonian of the model; the current operator $J(\bm{q})$ is determined by its 
kernel $\langle\bm{r}'|J(\bm{q})|\bm{r}\rangle=
\exp(i\bm{q}\cdot\bm{r})\,\delta^{(3)}(\bm{r}-\bm{r}')$. 
We do not take the spin of the current into account, therefore the basic quantum-mechanical Green function 
is the same for both types of form factors discussed above.

\subsection{Elastic form factor}
The elastic form factor of the ground state is given in terms of its wave function~$\Psi$~by
\begin{eqnarray}
\label{QM_ff}
F_{\rm el}(Q)=\langle\Psi|J(\bm{q})|\Psi\rangle =
\int{\rm d}^3r\,\exp({\rm i}\bm{q}\cdot\bm{r})\,|\Psi(\bm{r})|^2=\int{\rm d}^3k\Psi(\bm{k})
\,\Psi(\bm{k}+\bm{q}),\qquad Q\equiv|\bm{q}|. 
\end{eqnarray}
Here, $\Psi$ is the ground state of the Hamiltonian
\begin{equation}
\label{QM_H}H=\frac{\bm{k}^2}{2m}-\frac{\alpha}{r}+V_{\rm conf}(r),\qquad r\equiv|\bm{r}|.
\end{equation}
Because of the presence of the Coulomb interaction in the potential, the asymptotic behaviour of the form factor at 
large values of $Q$ is given by the factorization theorem \cite{brodsky}
\begin{equation}
\label{QM_factorization}
F_{\rm el}(Q)\xrightarrow[Q\to\infty]{}
\frac{16\pi\,\alpha\,m\,R_g}{Q^4},\qquad
R_g\equiv|\Psi(\bm{r}=\bm{0})|^2.
\end{equation}
The quantum-mechanical LD sum rule for the form factor $F_{\rm el}(Q)$ is rather similar to that~in~QCD: 
The double Borel transform ($E\to T$, $E'\to T'$) of (\ref{3pointQM}) may be written in the form  
\begin{eqnarray}
\label{3pointQM_Borel}
\Gamma^{\rm NR}(T,T',Q)=
\int dk'\,\exp\left(-\frac{k'^2}{2m}T'\right)
\int dk\,\exp\left(-\frac{k^2}{2m}T\right)
\,\Delta^{\rm NR}_{\rm pert}(k,k',Q)+\Gamma^{\rm NR}_{\rm power}(T,T',Q), 
\end{eqnarray}
where $\Gamma^{\rm NR}_{\rm power}(T,T',Q)$ describes the contribution of the confining interaction and 
$\Delta^{\rm NR}_{\rm pert}(k,k',Q)$ is 
the double spectral density of Feynman diagrams of nonrelativistic perturbation theory, see Appendix A for details. 

Setting $T'=T=0$ leads to the LD sum rule, in which case $\Gamma^{\rm NR}_{\rm power}$ vanishes \cite{lms1}. 
The low-energy region of perturbative diagrams---below some effective continuum threshold 
$k_{\rm eff}(Q)$---is assumed to be dual to the 
ground-state contribution, which reads $R_g\,F_{\rm el}(Q^2)$. Finally, we arrive at the following LD 
expression for the elastic form factor: 
\begin{eqnarray}
\label{2pt_QM}
R_g&=&\int_0^{k_{\rm eff}}{\rm d}k\,\rho^{\rm QM}_{\rm pert}(k)=\frac{k^3_{\rm eff}}{6\pi^2}+\alpha\,m\,\frac{k^2_{\rm eff}}{9\pi}+O(\alpha^2),
\\
\label{3pt_QM}
F^{\rm LD}_{\rm el}(Q)&=&\frac{1}{R_g}\int\limits_0^{k_{\rm eff}(Q)}\,{\rm d}k\int\limits_0^{k_{\rm eff}(Q)}\,{\rm d}k'\,
\Delta^{\rm QM}_{\rm pert}(k,k',Q).
\end{eqnarray}
The explicit result for $\Delta^{\rm QM}_{\rm pert}(k_1,k_2,Q)$ is given in Appendix A.

The factorization formula (\ref{QM_factorization}) is reproduced by the LD sum rule (\ref{3pt_QM}) 
if the momentum-dependent effective threshold behaves as  
\begin{equation}
\label{QM_LDa}
k_{\rm eff}(Q)\xrightarrow [Q\to\infty]{}k_{\rm LD}\equiv(6\pi^2\,R_g)^{1/3}.
\end{equation}


\subsection{Transition form factor}
The analogue of the $\pi\gamma$ transition form factor in quantum mechanics is given by
\begin{eqnarray}
\label{fnr}
F_{\rm trans}(Q, E)=\langle\Psi|J(\bm{q})\frac{1}{H-E}|r=0\rangle,
\end{eqnarray}
The case of one real and one virtual photon corresponds to $E=0$ and $Q\ne 0$. 
At large $Q$, the transition form factor $F_{\rm trans}(Q)\equiv F_{\rm trans}(Q,E=0)$ satisfies the factorization theorem 
\begin{equation}
\label{QM_factorization2}
F_{\rm trans}(Q)\xrightarrow[Q\to\infty]
{}
\frac{2m\sqrt{R_g}}{Q^2}.
\end{equation}
Recall that the behaviour (\ref{QM_factorization2}) does not require the Coulomb potential in the interaction 
and---in distinction to the factorization of the elastic form factor---emerges also for a purely confining interaction. 

The LD sum rule for the form factor $F_{\rm trans}(Q)$ is constructed on the basis of the same three-point function 
(\ref{3pointQM}) and has the form 
\begin{eqnarray}
\label{fnrdual}
F^{\rm LD}_{\rm trans}(Q)=\frac{1}{\sqrt{R_g}}\int\limits_0^{\bar k_{\rm eff}(Q)}\,{\rm d}k
\int\limits_0^{\infty}{\rm d}k'\,\Delta^{\rm QM}_{\rm pert}(k,k',Q).
\end{eqnarray}
Notice that the $k'$-integration is not restricted to the low-energy region since we do not isolate the ground-state 
contribution in the initial state. 
The asymptotical behaviour (\ref{QM_factorization2}) is correctly reproduced by Eq.~(\ref{fnrdual}) for 
\begin{equation}
\label{QM_LD2}
\bar k_{\rm eff}(Q\to\infty)=k_{\rm LD}.
\end{equation}


\subsection{Quantum-mechanical LD model}
As is obvious from (\ref{QM_LDa}) and (\ref{QM_LD2}), the effective thresholds for the elastic and for the 
transition form factors have the same limit at large $Q$: 
\begin{equation}
\label{QM_LD3}
k_{\rm eff}(Q\to\infty)=\bar k_{\rm eff}(Q\to\infty)=k_{\rm LD}.
\end{equation}
The LD model emerges when one {\it assumes} that also for intermediate $Q$ one may find a reasonable estimate 
for the form factors by setting 
\begin{equation}
\label{QM_LD}
k_{\rm eff}(Q)=\bar k_{\rm eff}(Q)=k_{\rm LD}.
\end{equation}
Similarly to QCD, the only property of the bound state which determines the form factor in the LD model is $R_g$.

\subsection{LD vs. exact effective threshold}
Let us now calculate the exact thresholds $k_{\rm eff}(Q)$ and $\bar k_{\rm eff}(Q)$ which reproduce the exact 
form factor by the LD expression; they are obtained by solving  
the LD sum rules (\ref{3pt_QM}) and (\ref{fnrdual}) using the exact form factors on the left-hand sides of these equations. 
The deviation of the LD threshold $k_{\rm LD}$ from these exact thresholds measures the error induced by the approximation 
(\ref{QM_LD}) and characterizes the accuracy of the LD model. 

For our numerical analysis we use parameter values relevant for hadron physics: 
$m=0.175$ GeV for the reduced constituent light-quark mass and $\alpha=0.3$. 
We considered several confining potentials 
\begin{equation}
\label{QM_conf}
V_{\rm conf}(r)=\sigma_n\,(m\,r)^n, \qquad n=2,1,1/2, 
\end{equation}
and adapt the strengths $\sigma_n$ in our confining interactions such that the Schr\"odinger equation yields for each 
potential the same value of the wave function at the origin, $\Psi({r}={0})=0.078$ GeV$^{3/2},$ which holds 
for $\sigma_2=0.71$ GeV, $\sigma_1=0.96$ GeV, and $\sigma_{1/2}=1.4$~GeV. 
The ground state then has a typical hadron size $\sim$ 1 fm. 

\begin{figure}[!ht]
\begin{center}
\begin{tabular}{cc}
\includegraphics[width=7.8cm]{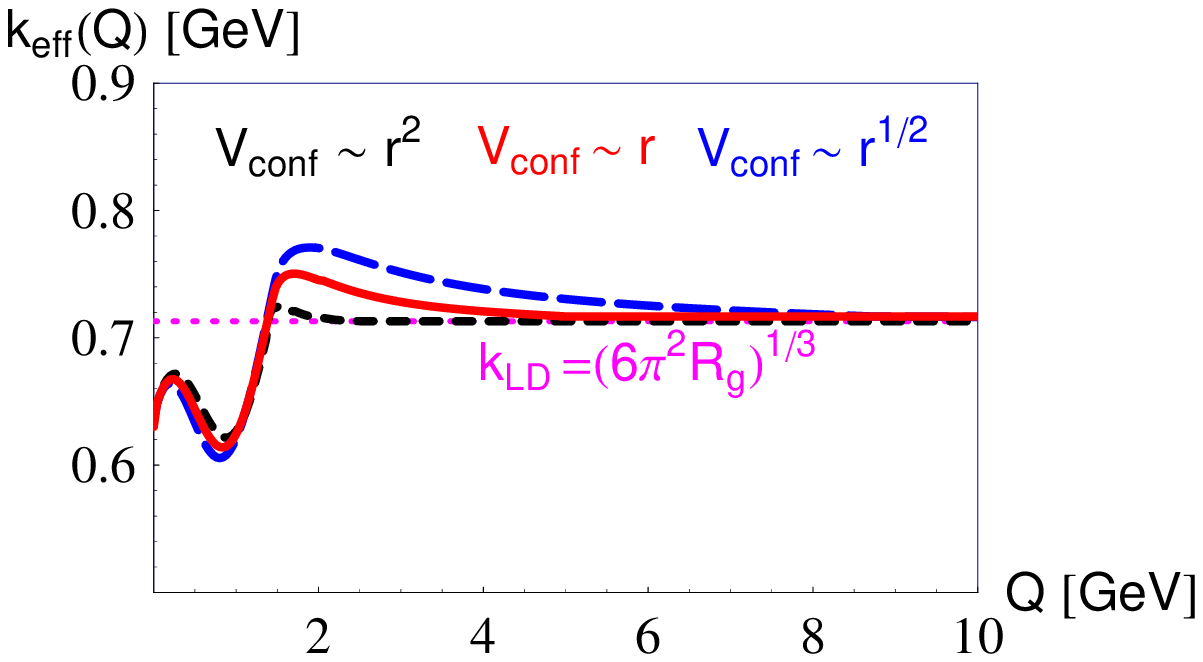}&
\includegraphics[width=7.8cm]{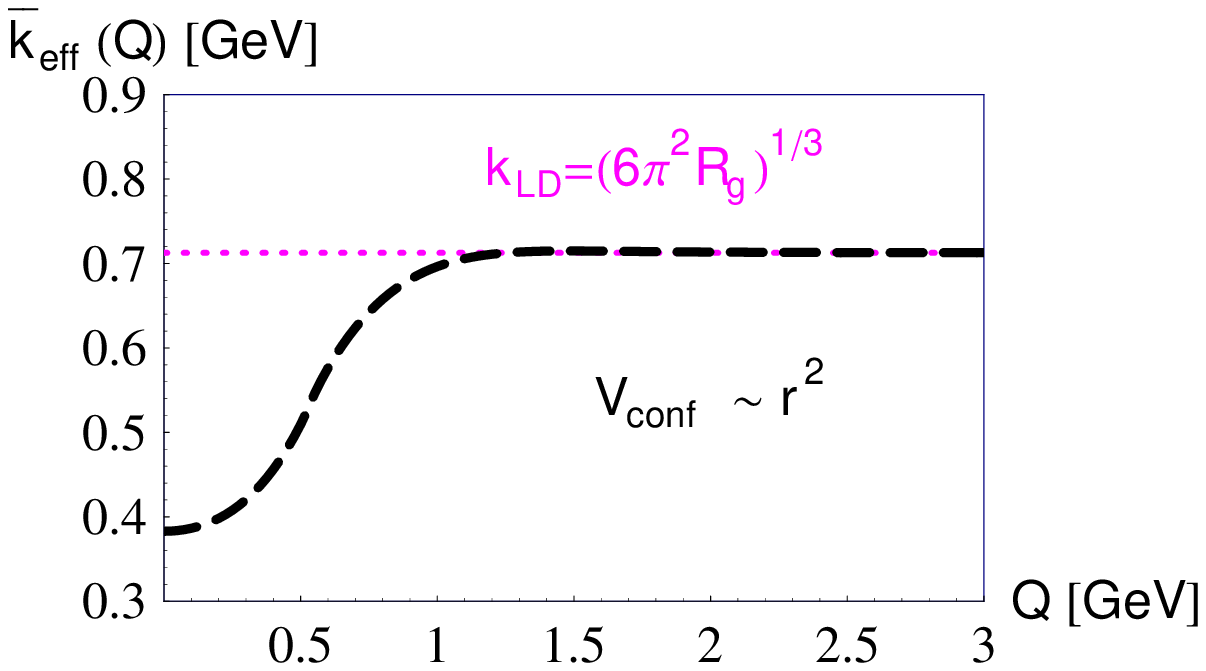}
\end{tabular}
\caption{\label{Plot:2}
Exact effective thresholds in quantum mechanics for the elastic (left) and the transition (right) form factors for 
different confining potentials. $R_{\rm g}\equiv |\Psi(r=0)|^2$.}
\end{center}
\end{figure}
Figure \ref{Plot:2} presents the exact effective thresholds. 
Independently of the details of the confining interaction, the accuracy of the LD approximation for 
the effective threshold and, respectively, the accuracy of the LD elastic form factor increases with 
$Q$ in the region $Q^2\ge 5-8$ GeV$^2$. 
For the transition form factor, the LD approximation works well starting with even smaller values of $Q$.


\section{The pion elastic form factor}
The $O(1)$ and $O(\alpha_s)$ spectral densities are at our disposal, and the only missing ingredient for obtaining 
the form factor is $s_{\rm eff}$. 
We know the rigorous constraints on the effective threshold---at $Q^2=0$ from the Ward identity 
and at $Q^2\to\infty$ from factorization---so it is easy to construct a model for
$s_{\rm eff}(Q^2)$ by a smooth interpolation between these values. A simple parameterization with a single 
constant $Q_0$ fixed by fitting the data at $Q^2=1$ GeV$^2$ might read \cite{blm2011jpg} 
\begin{equation}
\label{seff_LD}s_{\rm eff}(Q^2)=\frac{4\pi^2\,f^2_\pi}{1+\alpha_s(0)/\pi}
\left[1+\tanh\!\left(\frac{Q^2}{Q_0^2}\right)\frac{\alpha_s(0)}{\pi}\right],
\qquad Q_0^2=2.02\;{\rm GeV}^2.  
\end{equation}
According to Fig.~\ref{Plot:1}, our interpolation perfectly describes the well-measured data in the range 
$Q^2\approx0.5$--$2.5$~GeV$^2.$

Note that the effective continuum threshold $s_{\rm eff}(Q^2)$ in Eq.~(\ref{seff_LD}) approaches its 
limit $s_{\rm LD}$ already at $Q^2\approx4$--$5$ GeV$^2$. For $Q^2>4$--$5$ GeV$^2$, it practically 
coincides with the LD effective threshold of \cite{ld1}. Moreover, for $Q^2>5$--$6$ GeV$^2$~the
formula~(\ref{seff_LD}) is pretty close to the model of \cite{blm2008}. Obviously, the model labeled BLM in
Fig.~\ref{Plot:1} provides a perfect description of the available
$F_\pi(Q^2)$ data in the region $Q^2=1$--$2.5$ GeV$^2$. For $Q^2\ge3$--$4$ GeV$^2$, it reproduces well all the
data,~except~for~a point at $Q^2=10$ GeV$^2,$ where it is off the
present experimental value, which anyhow has a rather large error,
by~some two standard deviations.\footnote{It is virtually
impossible to construct models compatible with all experimental
results within $Q^2=2.5$--$10$ GeV$^2,$ as revealed by closer
inspection of Fig.~\ref{Plot:1}: those approaches which hit the
data at $Q^2=10$ GeV$^2$ overestimate the better-quality data
points~at~$Q^2\approx2$--$4$~GeV$^2$.}

Interestingly, in the region
$Q^2\ge3$--$4$ GeV$^2$ the BLM model yields considerably lower
predictions than the results of the different theoretical
approaches presented in Refs.~\cite{recent1,recent2,recent3,bakulev}. 

\begin{figure}[!hb]
\begin{center}
\begin{tabular}{cc}
\includegraphics[width=8.5cm]{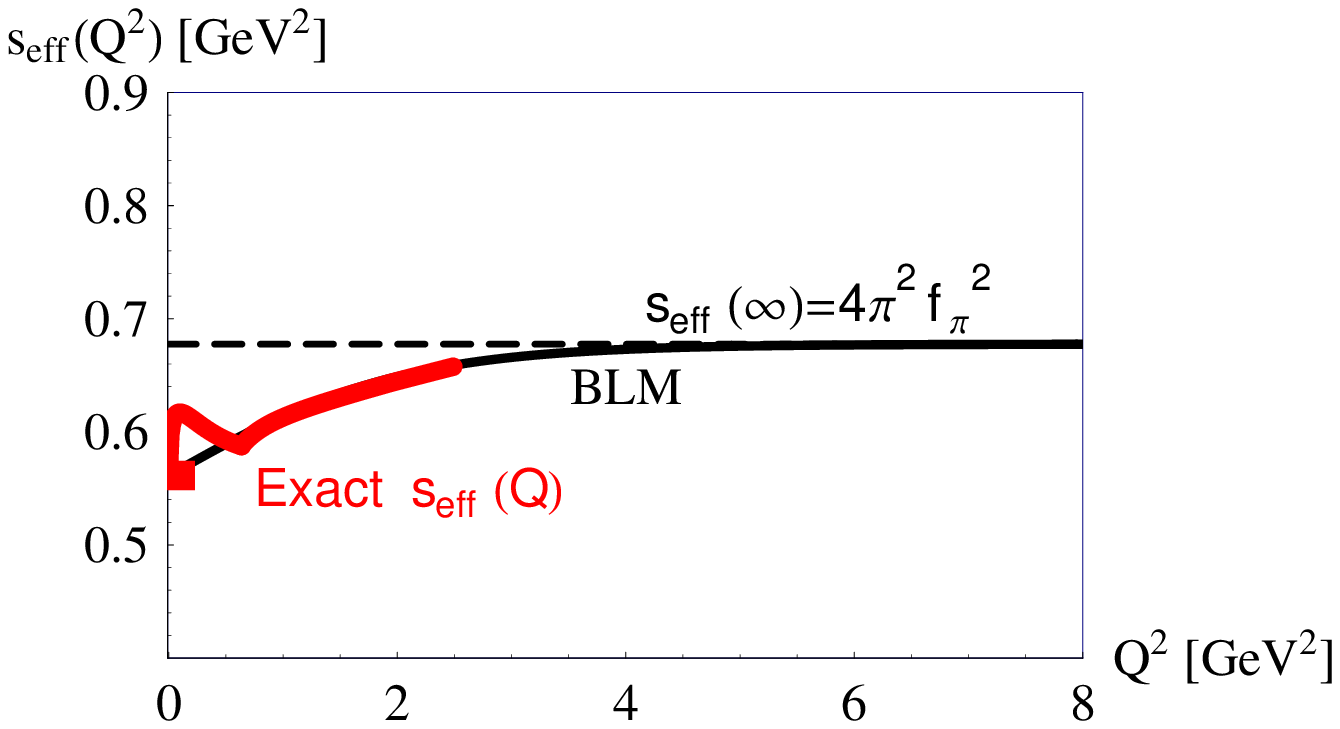}&
\includegraphics[width=8.5cm]{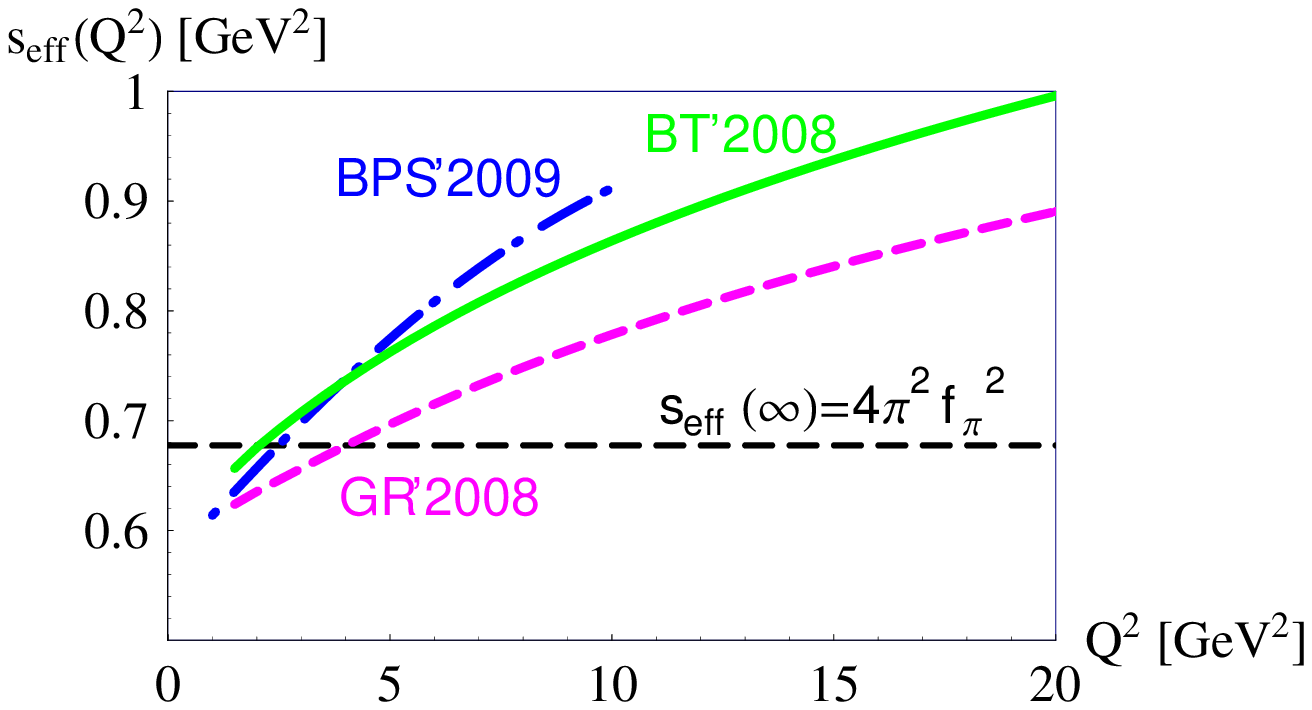}
\end{tabular}
\caption{\label{Plot:3}
Left: the ``equivalent effective threshold'' extracted from the data (red) vs. the improved LD model (BLM) 
of \cite{blm2011jpg}. Right: equivalent thresholds for the theoretical predictions displayed in Fig.~\protect\ref{Plot:1}. }
\end{center}
\end{figure}

For a given result for the pion form factor, we define the {\it equivalent effective threshold} as the quantity which 
reproduces this result by Eq.~(\ref{Fld1}). Figure~\ref{Plot:3} displays the equivalent effective thresholds recalculated 
from the data and from the theoretical predictions for the elastic form factor from Fig.~\ref{Plot:1}. 
For the $F_\pi(Q^2)$ predictions of \cite{recent1,recent2,bakulev}, the~corresponding
equivalent effective thresholds $s_{\rm eff}(Q^2)$ recalculated from (\ref{SR_G5}) are depicted in Fig.~\ref{Plot:3}: 
In all cases, they considerably exceed, for larger $Q^2,$ the LD
limit $s_{\rm LD}$ dictated by factorization. Moreover, their deviation from $s_{\rm LD}$ increases with~$Q^2$.

The exact effective threshold extracted from the accurate data at low $Q^2$ suggests that the LD limit may be 
reached already at relatively low values of $Q^2\approx 4-8$ GeV$^2$. However, the results in the right plot 
imply that the accuracy of the LD model still does not increase---or even decreases---with $Q^2$ even 
in the region $Q^2\simeq 20$ GeV$^2$, in conflict with both our experience from quantum mechanics and  
the hint from the data at low $Q^2$. 
We look forward to the future accurate data expected from JLab in the range up to $Q^2=8$ GeV$^2$. 

\section{The {\boldmath $(\pi^0,\eta,\eta')\to\gamma\gamma^*$} transition form factors}
\subsection{\boldmath $(\eta,\eta')\to\gamma\gamma^*$}
Before discussing the $\pi^0$ case, let us consider the $\eta$ and $\eta'$ decays. 
Here, one has to take properly into account the $\eta-\eta'$ mixing and the presence of two---strange and 
nonstrange---LD form factors. Following \cite{anisovich,feldmann,chernyak2009} we describe the flavor structure 
of $\eta$ and $\eta'$ as\footnote{For comparison with the form factors obtained in a scheme based on the 
octet--singlet mixing, we refer to \cite{teryaev2}.} 
\begin{eqnarray}
|\eta\rangle  = |\frac{\bar uu+\bar dd}{\sqrt{2}} \rangle\cos \phi-|\bar ss\rangle \sin \phi,\quad 
|\eta'\rangle = |\frac{\bar uu+\bar dd}{\sqrt{2}} \rangle\sin \phi+|\bar ss\rangle \cos \phi,\quad  \phi\approx 39.3^0. 
\end{eqnarray}
The $\eta$ and $\eta'$ form factors then take the form 
\begin{eqnarray}
\label{Feta}
F_{\eta\gamma}(Q^2)=
\frac{5}{3\sqrt2}F_{n\gamma}(Q^2) \cos \phi-\frac{1}{3} F_{s\gamma}(Q^2)\sin\phi,\quad
F_{\eta'\gamma}(Q^2)=
\frac{5}{3\sqrt2}F_{n\gamma}(Q^2)\sin \phi+\frac{1}{3}F_{s\gamma}(Q^2)\cos\phi.
\nonumber\\ 
\end{eqnarray}
Here, $F_{n\gamma}(Q^2)$ and $F_{s\gamma}(Q^2)$ are the form factors describing the transition of the nonstrange 
and $\bar ss$-components, respectively. 
The LD expressions for these quantities read 
\begin{eqnarray}
F_{n\gamma}(Q^2)=\frac{1}{f_n}\int\limits_0^{s_{\rm eff}^{(n)}(Q^2)}ds\,\sigma^{(n)}_{\rm pert}(s,Q^2),\quad 
F_{s\gamma}(Q^2)=\frac{1}{f_s}\int\limits_{4m_s^2}^{s_{\rm eff}^{(s)}(Q^2)}ds\,\sigma^{(s)}_{\rm pert}(s,Q^2), 
\end{eqnarray}
where $\sigma^{(n)}_{\rm pert}$ and $\sigma^{(s)}_{\rm pert}$ denote $\sigma_{\rm pert}$ with the corresponding 
quark propagating in the loop. Let us mention that for an isosinglet axial-vector current, a QCD axial anomaly 
contributes to the amplitude of interest \cite{anselm} and to the spectral densities of the form factors. 
This effect is of the order $\alpha_s^2$ and is not be expected to be important at large $Q^2$. 
An overall agreement of the form factor from the LD model for $\eta $ and $\eta'$ mesons with the data 
speaks in favour of this expectation. 

In numerical calculations we set $m_u=m_d=0$ and $m_s=100$ MeV.  
The LD model involves two separate effective thresholds for the nonstrange and 
the strange components \cite{feldmann}:
\begin{eqnarray}
s_{\rm eff}^{(n)}&=4\pi^2f_n^2,\qquad
f_n\approx1.07f_\pi,\qquad 
s_{\rm eff}^{(s)}&=4\pi^2f_s^2,\qquad f_s\approx1.36f_\pi.
\end{eqnarray}
According to the experience from quantum mechanics, the LD model may not perform well for small values of $Q^2$, 
where the true effective threshold is smaller than the LD threshold; however, for larger $Q^2$ the LD model in 
quantum mechanics gives accurate predictions for the form factors, as illustrated by~Fig.~\ref{Plot:3}. 
\begin{figure}[!hb]
\begin{center}
\begin{tabular}{cc}
\includegraphics[width=8.5cm]{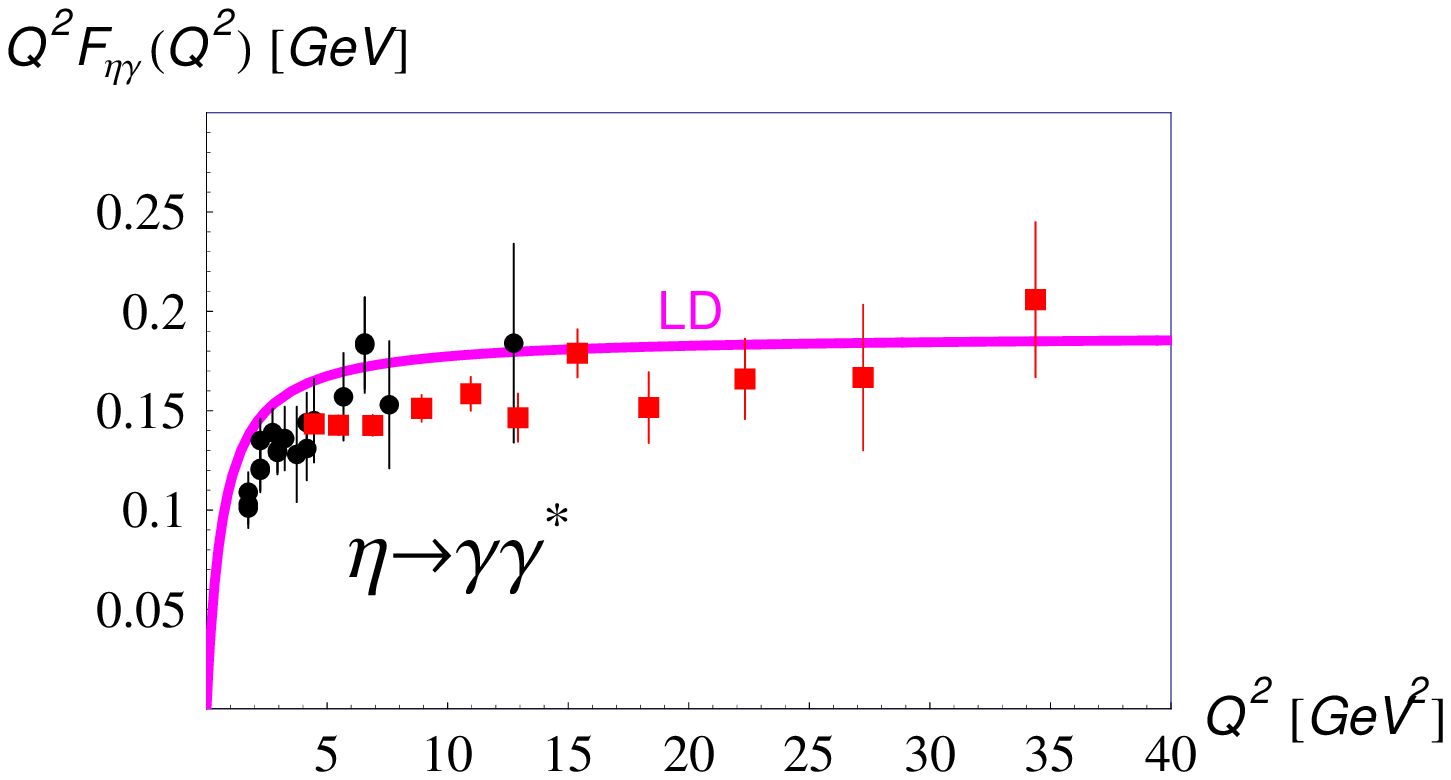} & 
\includegraphics[width=8.5cm]{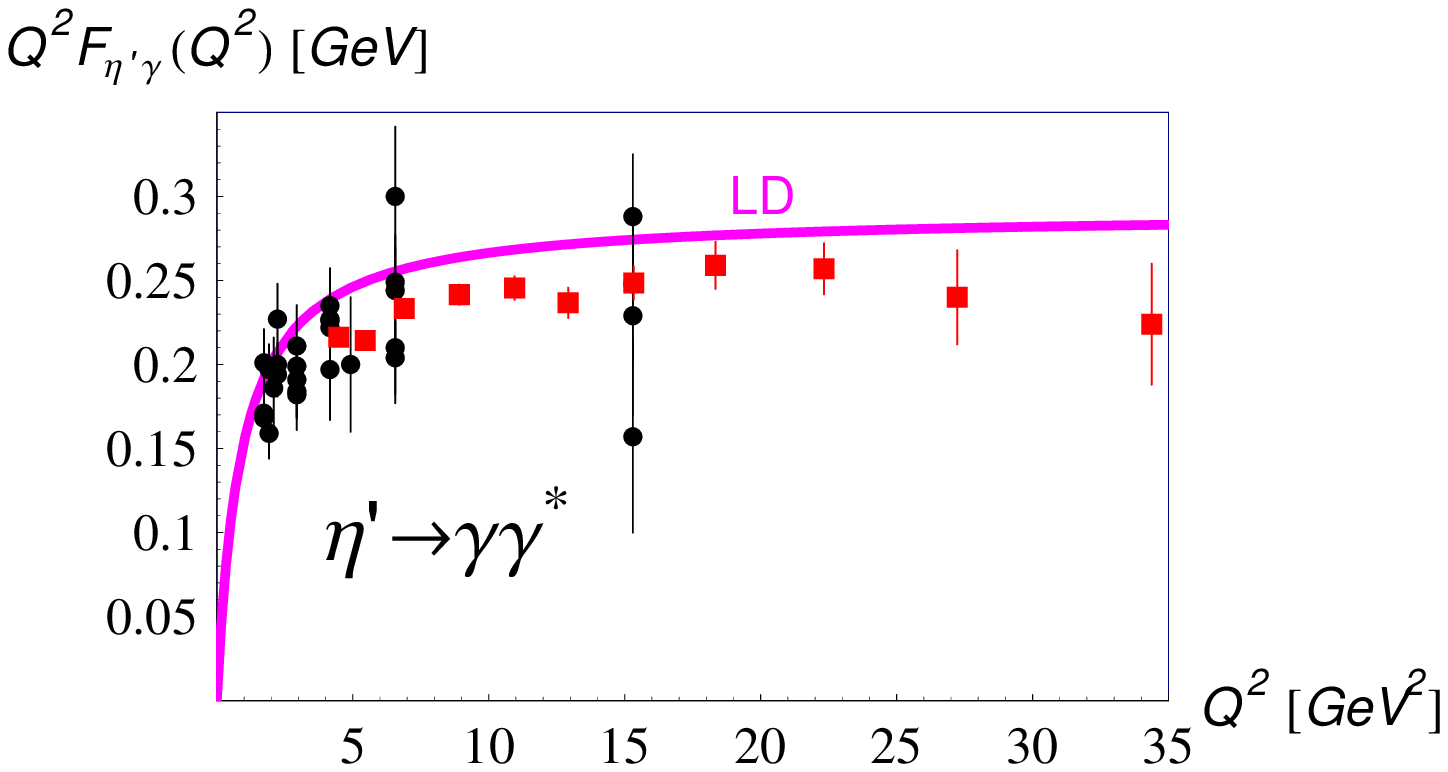}
\end{tabular}
\caption{\label{Plot:4}
LD predictions for $\eta$ and $\eta'$ vs. experimental data from \cite{cello-cleo} (black) and \cite{babar1} (red).}
\end{center}
\end{figure}
Figure~\ref{Plot:4} shows the corresponding predictions for $\eta$ and $\eta'$ mesons. One observes an overall 
agreement between the LD model and the data, meeting the expectation from quantum mechanics. 

\subsection{\boldmath $\pi^0\to\gamma\gamma^*$}
\begin{figure}[!hb]
\begin{center}
\begin{tabular}{cc}
\includegraphics[width=8.5cm]{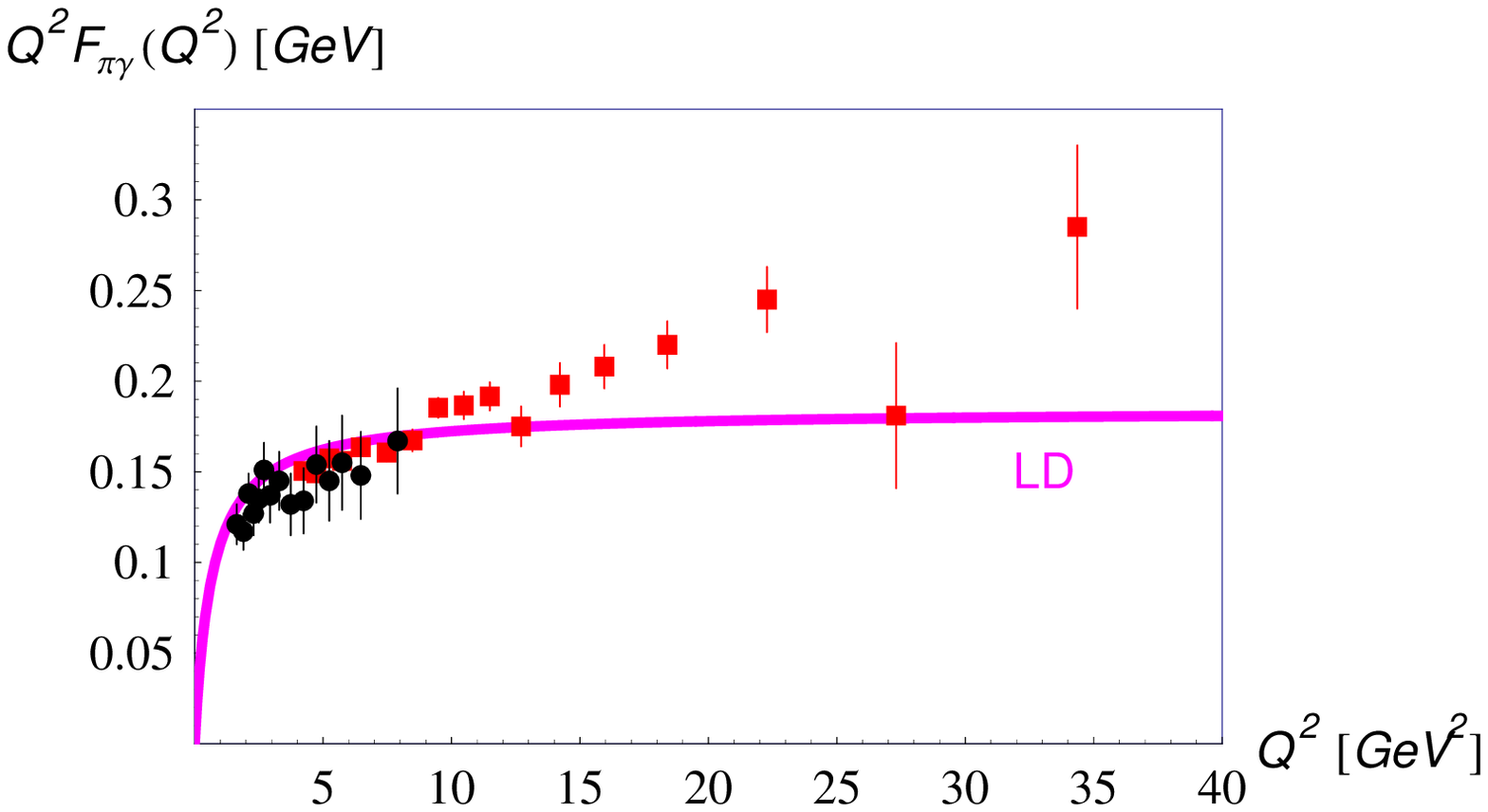}&
\includegraphics[width=8.5cm]{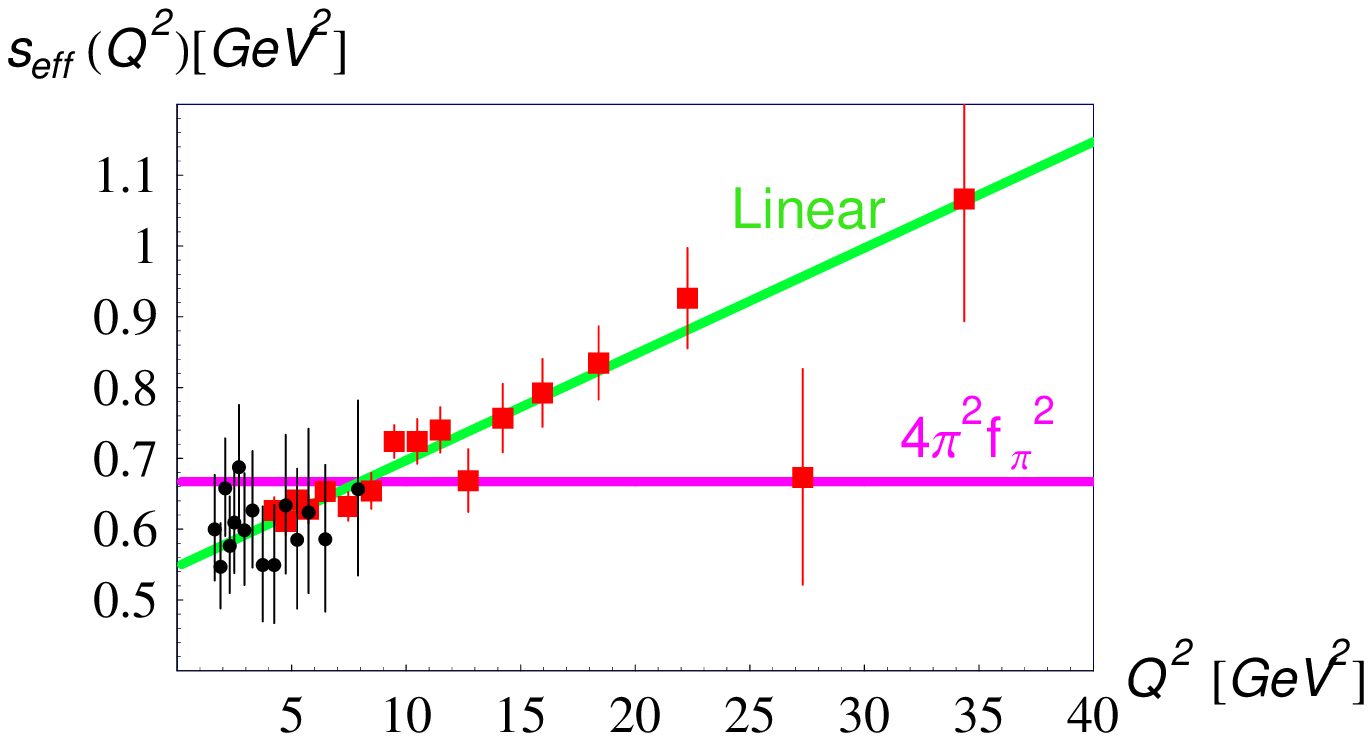}
\end{tabular}
\caption{\label{Plot:5}
The LD $\pi\gamma$ form factor vs. data from \cite{cello-cleo} (black) and  \cite{babar} (red) and the corresponding 
equivalent effective threshold.}
\end{center}
\end{figure}

Surprisingly, for the pion transition form factor, Fig.~\ref{Plot:5}, one observes a clear disagreement between 
the results from the LD model and the {\sc BaBar} data \cite{babar}. 
Moreover---in evident conflict with the $\eta$ and $\eta'$ results and the experience from quantum mechanics---the 
data implies that the violations of LD increase with $Q^2$ even in the region $Q^2\approx 40$ GeV$^2$! 
The effective threshold extracted from the {\sc BaBar} data is compatible with a linear growing function of $Q^2$ with no 
sign of approaching the LD limit. 

The $P\to\gamma$ transition form factors have been addressed recently in many publications (see, e.g.,
\cite{roberts,mikhailov,agaev,bt,dorokhov,kroll,teryaev2,blm2011prd,lcqm,czyz,pospelov,ms2012}).  
However, no convincing explanation of the full picture of the $P\to\gamma$ transition form factors has been offered. 
In fact, it is hard to find a convincing answer to the question why nonstrange components in $\eta$, $\eta'$, on the one hand, 
and in $\pi^0$, on the other hand, should behave so much differently?

\section{Summary}
We presented the analysis of the pion elastic and the $\pi^0$, $\eta$, $\eta'$ transition form factors from the LD 
version of QCD sum rules. 
The main emphasis was laid on the attempt to probe the accuracy of this approximate method and the reliability of its predictions. 
%
Our main conclusions are as follows:

\begin{itemize}
\item
\noindent 
{\bf The elastic form factor}: 
Our quantum-mechanical analysis suggests that the LD model should work 
increasingly well in the region $Q^2\ge 4-8$ GeV$^2$, independently of the details of the confining interaction. 
For arbitrary confining interaction, the LD model gives very accurate results for $Q^2\ge 20-30$ GeV$^2$. 
The accurate data on the pion form factor at small momentum transfers indicate that the LD limit for the effective 
threshold, $s^{\rm LD}_{\rm eff}=4\pi^2 f_\pi^2$, may be reached already at relatively low values $Q^2=5-6$ GeV$^2$; 
thus, large deviations from the LD limit at $Q^2=20-50$~GeV$^2$ reported in some recent publications 
\cite{recent1,recent2,recent3,bakulev} appear to us rather unlikely. 
\item
\noindent 
{\bf The {\boldmath $P\to\gamma\gamma^*$} transition form factor}: 
We conclude from the quantum-mechanical analysis that 
the LD model should work well in the region $Q^2\ge$ a few GeV$^2$. 
Indeed, for the $\eta\to\gamma\gamma^*$ and $\eta'\to\gamma\gamma^*$ form factors, the 
predictions from LD model in QCD work reasonably well. 
Surprisingly, for the $\pi\to \gamma\gamma^*$ form factor the present {\sc BaBar} data indicate an increasing violation 
of local duality, corresponding to 
a linearly rising effective threshold, even at $Q^2$ as large as 40 GeV$^2$. 
This puzzle has so far no compelling theoretical explanation. 
Our conclusion agrees with the findings of \cite{roberts,mikhailov,bt} obtained 
from other theoretical approaches.
\end{itemize}
\vspace{.5cm}

\noindent 
{\bf Acknowledgments.} 
We are grateful to S.~Mikhailov, O.~Nachtmann, S.~Simula, O.~Teryaev, and particularly to B.~Stech for valuable discussions. 
D.~M.\ was supported by the Austrian Science Fund (FWF) under Project No.~P22843. 
This work was supported in part by grants for leading scientific schools 1456.2008.2 and 3920.2012.2, 
and by FASI State Contract No.~02.740.11.0244.


\appendix\section{\label{Sec:App}Perturbative expansion of Green functions in quantum mechanics}
We construct the perturbative
expansions of both polarization operator and vertex function in
quantum mechanics.

\subsection{Polarization operator}
\begin{figure}[hb!]
\begin{center}
\includegraphics[width=11cm]{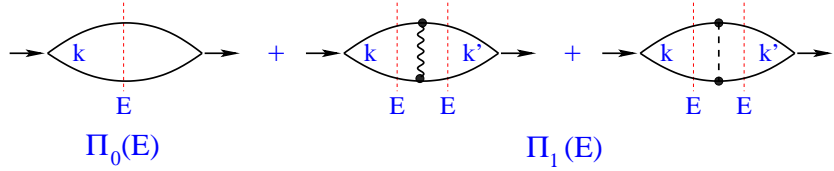}
\caption{\label{Fig:1}Expansion of the polarization operator in
terms of Coulomb (wavy line) and confining (dashed line)
interaction potentials.}
\end{center}
\end{figure}

The polarization operator
$\Pi(E)$ is defined by \cite{qmsr}
\begin{eqnarray}
\Pi(E)=\langle\bm{r}'=\bm{0}|G(E)|\bm{r}=\bm{0}\rangle,
\end{eqnarray}
where $G(E)$ is the full Green function, i.e., $G(E)=(H-E)^{-1},$
defined by the model Hamiltonian under consideration
\begin{eqnarray}
H=H_0+V(r),\qquad H_0\equiv\frac{\bm{k}^2}{2m},\qquad
r\equiv|\bm{r}|.
\end{eqnarray}
The expansion of the full Green
function $G(E)$ in powers of the interaction potential $V$ has the
well-known form
\begin{eqnarray}
\label{lippmann}
G(E)=G_0(E)-G_0(E)VG_0(E)+G_0(E)VG_0(E)VG_0(E)+\cdots,
\end{eqnarray}
with $G_0(E)=(H_0-E)^{-1}.$ 
It generates the corresponding expansion of $\Pi(E)$:
\begin{eqnarray}
\label{seriesPi}
\Pi(E)=\Pi_0(E)+\Pi_1(E)+\cdots.
\end{eqnarray}
Explicitly, one
finds\begin{align}\Pi_0(E)&=\frac{1}{(2\pi)^3}\int\frac{{\rm
d}^3k}{\frac{\bm{k}^2}{2m}-E},\\
\Pi_1(E)&=-\frac{1}{(2\pi)^6}
\int\frac{{\rm d}^3k}{\frac{\bm{k}^2}{2m}-E}\frac{{\rm
d}^3k'}{\frac{\bm{k}'^2}{2m}-E}
V\!\left((\bm{k}-\bm{k}')^2\right).
\end{align}
We consider
interaction potentials $V(r)$ which consist of a Coulombic and a
confining part:
\begin{eqnarray}
V(r)=-\frac{\alpha}{r}+V_{\rm conf}(r).
\end{eqnarray}
Then the expansion (\ref{seriesPi}) becomes
a double expansion in powers of the Coulomb coupling $\alpha$ and
the confining potential $V_{\rm conf}$ (see Fig.~\ref{Fig:1}).

The contribution to $\Pi(E)$ arising from the Coulombic potential
is referred to as the perturbative contribution,~$\Pi_{\rm pert}$.
The contributions involving the confining potential $V_{\rm conf}$
(including the mixed terms receiving contributions from both
confining and Coulomb parts) are referred to as the power
corrections, $\Pi_{\rm power}.$ For instance, the first-order
perturbative contribution reads
\begin{eqnarray}
\Pi_1^{(\alpha)}(E)
=\frac{1}{(2\pi)^6}\int\frac{{\rm d}^3k}{\frac{\bm{k}^2}{2m}-E}
\frac{{\rm d}^3k'}{\frac{\bm{k}'^2}{2m}-E}
\frac{4\pi\alpha}{(\bm{k}-\bm{k}')^2}=\frac{\alpha m}{8\pi^2}
\int\frac{{\rm d}^3k}{\left(\frac{\bm{k}^2}{2m}-E\right)|\bm{k}|}.
\end{eqnarray}
The integral diverges but becomes convergent after applying the Borel
transformation $1/(a-E)\to\exp(-aT)$.\footnote{Note that the Borel
transform of the Green function $(H-E)^{-1}$ yields the
quantum-mechanical time-evolution operator in imaginary~time
$U(T)=\exp(-HT)$.} The Borel-transformed polarization operator
$\Pi(T)$ has the form \cite{voloshin95}
\begin{eqnarray}
\Pi(T)=\Pi_{\rm pert}(T)+\Pi_{\rm power}(T),\qquad 
\Pi_{\rm pert}(T)=\left(\frac{m}{2\pi T}\right)^{3/2} 
\left[1+\sqrt{2\pi mT}\alpha+\frac{1}{3}m\pi^2T\alpha^2+O(\alpha^3)\right].
\end{eqnarray}
In the LD limit, that is, for $T\to0$, only $\Pi_{\rm pert}(T)$ will be relevant. Nevertheless,
as an illustration we provide also the result for the power corrections $\Pi_{\rm power}(T)$ for the case of a
harmonic-oscillator confining potential $V_{\rm conf}(r)=m\omega^2r^2/2$:
\begin{eqnarray}
\Pi_{\rm power}(T)=\left(\frac{m}{2\pi T}\right)^{3/2}
\left[-\frac{1}{4}\omega^2T^2\left(1+\frac{11}{12}\sqrt{2\pi mT}\alpha\right)+\frac{19}{480}\omega^4T^4\right].
\end{eqnarray}
Let us point out that $\Pi_{\rm power}(T=0)$ vanishes, similar to QCD. The radiative corrections in $\Pi_{\rm pert}(T)$ have a
less~singular behaviour compared to the free Green function, so the system behaves as quasi-free system. In QCD such a behaviour,
frequently regarded as an indication of asymptotic freedom, occurs due to the running of the strong coupling $\alpha_s$~and~its
vanishing at small distances. Interestingly, in the nonrelativistic potential model this feature is built-in automatically.

Now, according to the standard procedures of the method of sum
rules, the dual correlator is obtained by applying~a low-energy
cut at some threshold $\bar k_{\rm eff}$ in the spectral
representation for the perturbative contribution to the
correlator:
\begin{eqnarray}
\Pi_{\rm dual}(T,\bar k_{\rm eff})=
\frac{1}{2\pi^2}\int\limits_0^{\bar k_{\rm eff}}{\rm d}k\,k^2
\exp\!\left(-\frac{k^2}{2m}T\right)\!\left[1+\frac{\pi m\alpha}{k}
+\frac{(\pi m\alpha)^2}{3k^2}+O(\alpha^3)\right]+\Pi_{\rm power}(T).
\end{eqnarray}
By construction, the dual correlator
$\Pi_{\rm dual}(T,\bar k_{\rm eff})$ is related to the
ground-state contribution by
\begin{eqnarray}
\label{Pidual}
\Pi_{\rm dual}(T,\bar k_{\rm eff})=\Pi_{\rm g}(T)\equiv R_{\rm
g}\exp(-E_{\rm g}T),\qquad R_{\rm g}\equiv|\psi_{\rm g}(r=0)|^2.
\end{eqnarray}
As we have shown in our previous studies
of potential models, the effective continuum threshold defined
according to (\ref{Pidual}) is a function of the Borel time
parameter $T$. For $T=0,$ one finds
\begin{eqnarray}
\label{picut}
\Pi_{\rm dual}(\bar k_{\rm eff},T=0)=\frac{1}{6\pi^2}\,{\bar k_{\rm eff}}^3+\frac{\alpha m}{4\pi}\,\bar k_{\rm eff}^2+\cdots.
\end{eqnarray}

\subsection{Vertex function}
\begin{figure}[hb!]
\begin{center}
\includegraphics[width=12cm]{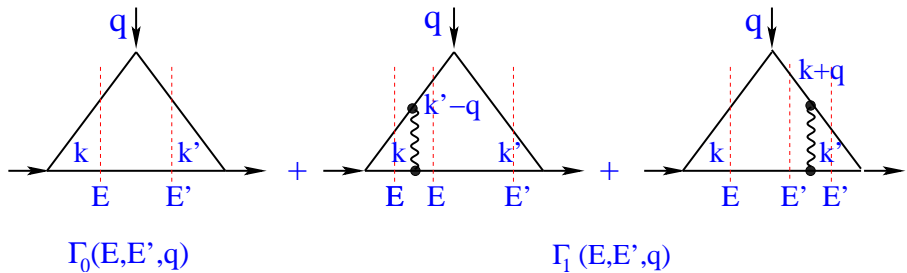}
\caption{\label{Fig:7}Nonrelativistic Feynman diagrams
representing the lowest perturbative contributions to the vertex
function $\Gamma(E,E',Q)$.}
\end{center}
\end{figure}

We now calculate the vertex function $\Gamma(E,E',Q),$ defined by
\begin{eqnarray}
\label{Gamma3pt}
\Gamma(E,E',Q)=\langle\bm{r}'=\bm{0}|G(E)J(\bm{q})G(E')|\bm{r}=\bm{0}\rangle,
\qquad Q\equiv|\bm{q}|,
\end{eqnarray}
where $J(\bm{q})$ is the
operator which adds a momentum $\bm{q}$ to the interacting
constituent. The expansions (\ref{lippmann}) of the~full Green
functions $G(E)$ and $G(E')$ in powers of the interaction entail a
corresponding expansion of
$\Gamma(E,E',Q),$~cf.~Fig.~\ref{Fig:7}:
\begin{eqnarray}
\label{Eq:vertexp}
\Gamma(E,E',Q)=\Gamma_0(E,E',Q)+\Gamma_1(E,E',Q)+\cdots.
\end{eqnarray}

For the vertex functions $\Gamma_i(E,E',Q),$ $i=0,1,\dots,$ in
(\ref{Eq:vertexp}), their double spectral representations may be
written~as
\begin{eqnarray}
\label{Gamma_double}
\Gamma_i(E,E',Q)=\int\frac{{\rm d}z}{\frac{z}{2m}-E}\frac{{\rm
d}z'}{\frac{z'}{2m}-E'}\,\Delta_i(z,z',Q).
\end{eqnarray}
The vertex functions $\Gamma_i(E,E',Q=0)$ and the polarization operators
$\Pi_i(E)$ satisfy the Ward identities
\begin{eqnarray}
\label{WI}
\Gamma_i(E,E',Q=0)=\frac{\Pi_i(E)-\Pi_i(E')}{E-E'},
\end{eqnarray}
which are equivalent to the following relations between the
corresponding spectral densities:
\begin{eqnarray}
\label{WI1}
\lim\limits_{Q\to0}\Delta_i(z,z',Q)=\delta(z-z')\,\rho_i(z).
\end{eqnarray}
We shall consider the double Borel transform $E\to T$ and $E'\to
T'$: $(a-E)^{-1}\to\exp(-aT),$ $(a'-E')^{-1}\to\exp(-a'T')$.
Equation (\ref{WI1}) leads to the following Ward identities for
the Borel images:
\begin{eqnarray}
\label{WI2}
\Gamma_i(T,T',Q=0)=\Pi_i(T+T').
\end{eqnarray}

\subsubsection{One-loop contribution $\Gamma_0$ to the vertex function}
The zero-order one-loop term has the form
\begin{eqnarray}
\Gamma_0(E,E',Q) =\frac{1}{(2\pi)^3}\int\frac{{\rm
d}^3k}{\left(\frac{\bm{ k}^2}{2m}-E\right)\!
\left(\frac{(\bm{k}+\bm{q})^2}{2m}-E'\right)}.
\end{eqnarray}
This may be written as the double spectral representation
\begin{eqnarray}
\Gamma_0(E,E',Q)=\int\frac{{\rm d}z}{\frac{z}{2m}-E}\frac{{\rm
d}z'}{\frac{z'}{2m}-E'}\,\Delta_0(z,z',Q),
\end{eqnarray}
where
\begin{eqnarray}
\Delta_0(z,z',Q)=\frac{1}{(2\pi)^3}\int{\rm d}^3k\,
\delta\!\left(z-\bm{k}^2\right)\delta\!\left(z'-(\bm{k}-\bm{q})^2\right)
=\frac{1}{(2\pi)^3}\frac{\pi}{2Q}\,\theta\!\left((z'-z-Q^2)^2-4zQ^2<0\right).
\end{eqnarray}
Hereafter, we use the notations $k\equiv\sqrt{z}$ and
$k'\equiv\sqrt{z'}$. In terms of the variables $k$ and $k',$ the
$\theta$ function takes the~form
\begin{eqnarray}
\theta\!\left((z'-z-Q^2)^2-4zQ^2<0\right)
=\theta\!\left(|k-Q|<k'<k+Q\right).
\end{eqnarray}

\subsubsection{Two-loop contribution $\Gamma_1$ to the vertex function}
We consider here only corrections related to the {\em Coulomb\/} potential, 
since power corrections induced by the
confining interaction vanish in the LD limit. The two-loop
$O(\alpha)$ correction receives two contributions and has the form
\begin{eqnarray}
\Gamma_1(E,E',Q)=
\frac{1}{(2\pi)^6}\int\frac{{\rm d}^3k}{\frac{\bm{k}^2}{2m}-E}
\frac{{\rm d}^3k'}{\frac{\bm{k}'^2}{2m}-E'}
\frac{4\pi\alpha}{\left(\bm{k}-(\bm{k}'-\bm{q})\right)^2}
\left[\frac{1}{\frac{(\bm{k}'-\bm{q})^2}{2m}-E'}
+\frac{1}{\frac{(\bm{k}+\bm{q})^2}{2m}-E}\right].
\end{eqnarray}
Having in mind the subsequent application of a double
Borel transformation in $E$ and $E',$ it is convenient to
represent $\Gamma_1$ as a sum of two terms,
$\Gamma_1=\Gamma_1^{(a)}+\Gamma_1^{(b)}$, with
\begin{align}
\Gamma_1^{(a)}(E,E',Q)&=\frac{\alpha m}{8\pi^5}\int\frac{{\rm
d}^3k'}{\left(\frac{\bm{k}'^2}{2m}-E'\right)
\left(\frac{(\bm{k}'-\bm{q})^2}{2m}-E\right)}\int\frac{{\rm
d}^3k}{\left(\bm{k}-(\bm{k}'-\bm{q})\right)^2
\left[\bm{k}^2-(\bm{k}'-\bm{q})^2\right]}\nonumber\\
&+\frac{\alpha m}{8\pi^5}\int\frac{{\rm
d}^3k}{\left(\frac{\bm{k}^2}{2m}-E\right)
\left(\frac{(\bm{k}+\bm{q})^2}{2m}-E'\right)}\int\frac{{\rm
d}^3k'}{\left((\bm{k}+\bm{q})-\bm{k}'\right)^2
\left[\bm{k}'^2-(\bm{k}+\bm{q})^2\right]},\nonumber\\
\label{A.25}
\Gamma_1^{(b)}(E,E',Q)&=\frac{\alpha m}{8\pi^5}\int\frac{{\rm
d}^3k}{\frac{\bm{k}^2}{2m}-E}\frac{{\rm
d}^3k'}{\frac{\bm{k}'^2}{2m}-E'}\frac{1}{(\bm{k}+\bm{q}-\bm{k}')^2}
\left[\frac{1}{(\bm{k}'-\bm{q})^2-\bm{k}^2}
+\frac{1}{(\bm{k}+\bm{q})^2-\bm{k}'^2}\right].
\end{align}
The double Borel transformation in $E\to T$ and $E'\to T'$ is now
easily performed.

Let us start with $\Gamma_1^{(a)}.$ One integration in
$\Gamma_1^{(a)}$ may be performed, leading to
\begin{eqnarray}
\Gamma_1^{(a)}(E,E',Q)=\frac{\alpha m}{16\pi^2}\left[\int
\frac{{\rm d}^3k'}{\left(\frac{(\bm{k}'+\bm{q})^2}{2m}-E'\right)\!
\left(\frac{\bm{k}'^2}{2m}-E\right)|\bm{k}'|}+\int\frac{{\rm
d}^3k}{\left(\frac{(\bm{k}-\bm{q})^2}{2m}-E\right)\!
\left(\frac{\bm{k}^2}{2m}-E'\right)|\bm{k}|}\right].
\end{eqnarray}
The first term corresponds to the contribution of the ``left''
two-loop diagram in Fig.~\ref{Fig:7}, i.e., with the
potential~before~the interaction with the current $J(\bm{q}),$
while the second term is represented by the ``right'' two-loop
diagram in Fig.~\ref{Fig:7}.~~The corresponding double spectral
densities have a form very similar to $\Delta_0$:
\begin{equation}
\Delta_{1L}^{(a)}(k,k',Q)=\frac{\alpha m}{16\pi}\frac{\pi}{2Q}
\frac{1}{k}\,\theta\!\left(|k-Q|<k'<k+Q\right)\theta(0<k)\,\theta(0<k'),
\quad\Delta_{1R}^{(a)}(k,k',Q)=\Delta_{1L}^{(a)}(k',k,Q).
\end{equation}

Explicit calculations yield the following double spectral
densities of the two contributions to $\Gamma_1^{(b)}$
related~to~the~``left'' and ``right'' two-loop diagrams in
Fig.~\ref{Fig:7}:
\begin{equation}
\Delta_{1L}^{(b)}(k,k',Q)=\frac{\alpha m}{32\pi^6}
\frac{1}{Qk}\left[\log^2\!\left(\left|\frac{k'-Q+k}{k'-Q-k}\right|\right)
-\log^2\!\left(\left|\frac{k'+Q+k}{k'+Q-k}\right|\right)\right],\quad
\Delta_{1R}^{(b)}(k,k',Q)=\Delta_{1L}^{(b)}(k',k,Q).
\end{equation}
At $Q=0$, $\Gamma_1^{(a)}(T,T',Q=0)$ satisfies the Ward identity,
$\Gamma_1^{(a)}(T,T',Q=0)=\Pi_1^{(\alpha)}(T+T')$, whereas
$\Gamma_1^{(b)}(T,T',Q=0)$ vanishes: $\Gamma_1^{(b)}(T,T',Q=0)=0.$
For large $Q$ and $T,T'\ne0$, $\Gamma_1^{(b)}(T,T',Q)$ assumes a
factorizable form (see Eq.~(\ref{A.25})):
\begin{eqnarray}
\Gamma_1^{(b)}(T,T',Q)\to\frac{16\pi\alpha
m}{Q^4}\,\Pi_0(T)\,\Pi_0(T').
\end{eqnarray}
At the same time, both
$\Gamma_0(T,T',Q)$ and $\Gamma_1^{(a)}(T,T',Q)$ are exponentially
suppressed for large $Q$ and $T,T'\ne0$. Hence, $\Gamma_1^{(b)}$
determines the large-$Q$ behaviour of the vertex function.

\subsubsection{Dual correlator}
The dual correlator $\Gamma_{\rm dual}(T,T',Q)$ is constructed in a standard way, 
by application of a low-energy cut to the~double spectral representation of the
perturbative contribution to (\ref{Gamma_double}):
\begin{eqnarray}
\label{GammaBorel_double}\Gamma_{\rm dual}(T,T',Q)=
\int\limits_0^{k_{\rm eff}(Q,T)}{\rm d}k\,2k
\exp\!\left(-\frac{k^2}{2m}\,T\right)\int\limits_0^{k_{\rm
eff}(Q,T')}{\rm d}k'\,2k'\exp\!\left(-\frac{k'^2}{2m}\,T'\right)
\Delta(z,z',Q)+\Gamma_{\rm power}(T,T',Q).\qquad
\end{eqnarray}
By construction, the dual correlator corresponds to the ground-state
contribution $\exp(-E_g T)\exp(-E_g T')\,R_g\,F_g(Q).$

In the LD limit $T=0$ and $T'=0,$ $\Gamma_{\rm power}(T,T',Q)$
vanishes and the ground-state form factor $F_g(Q)$ is
related~to~the low-energy part of the perturbative contribution
considered above:
\begin{eqnarray}
\label{LD3ptsr}
\int\limits_0^{k_{\rm eff}(Q)}{\rm d}k\,2k\int\limits_0^{k_{\rm
eff}(Q)}{\rm d}k'\,2k'\Delta(k,k',Q)=F_{g}(Q)\,R_{g},
\end{eqnarray}
with
$\Delta(k,k',Q)=\Delta_0(k,k',Q)+\Delta_{1L}^{(a)}(k,k',Q)
+\Delta_{1R}^{(a)}(k,k',Q)+\Delta_{1L}^{(b)}(k,k',Q)
+\Delta_{1R}^{(b)}(k,k',Q)+O(\alpha^2)$.

In order to provide the correct normalization $F_g(Q=0)=1$ of the elastic form factor $F_g(Q),$ 
the effective thresholds should be related to each other according to $k_{\rm eff}(Q=0)=\bar k_{\rm eff}$; 
then the form factor is correctly normalized due to~the
Ward identity (\ref{WI1}) satisfied by the spectral densities.


\end{document}